\documentclass[preprint]{elsarticle}
\usepackage{amssymb}
\usepackage{amsfonts}
\usepackage{amsmath,amsthm}
\usepackage{graphicx}
\usepackage{graphics}
\usepackage{array}
\usepackage{subfigure}
\usepackage{rotating}
\journal{\textbf{.}}
\oddsidemargin 0cm \numberwithin{equation}{section}

\newcommand{\showx}{x}
\begin{document}
\title{
Numerical pricing of American options under two stochastic factor models with jumps using a meshless local Petrov-Galerkin method
%
}
\author{Jamal Amani Rad}
\ead{j.amanirad@gmail.com;j\_amanirad@sbu.ac.ir}
\author{Kourosh Parand\corref{cor1}}
\ead{k\_parand@sbu.ac.ir}
\cortext[cor1]{Corresponding author}
\address{Department of
Computer Sciences, Faculty of Mathematical Sciences, Shahid Beheshti University, Evin, P.O. Box 198396-3113,Tehran,Iran}
\begin{abstract}
The most recent update of financial option models is American options under stochastic volatility models with jumps in returns (SVJ) and stochastic volatility
models with jumps in returns and volatility (SVCJ).
To evaluate these options, mesh-based methods are applied in a number of papers but it is well-known that these methods depend strongly on the mesh properties which is the major disadvantage of them.
Therefore, we propose the use of the meshless methods to solve the aforementioned options models, especially in this work we select and analyze
one scheme of them, named local radial point interpolation (LRPI) based on Wendland's compactly supported radial basis functions (WCS-RBFs) with $C^6$
, $C^4$ and $C^2$ smoothness degrees.
The LRPI method which is a special type of meshless local Petrov-Galerkin method (MLPG), offers several advantages over the
mesh-based methods, nevertheless it has never been applied to option
pricing, at least to the very best of our knowledge.
These schemes are the truly meshless methods, because, a traditional non-overlapping continuous mesh
is not required, neither for the construction of the shape functions, nor for
the integration of the local sub-domains.
In this work, the American option which is a free boundary problem, is reduced to a problem with fixed boundary using a Richardson extrapolation technique.
Then the implicit-explicit (IMEX) time stepping scheme is employed for the time derivative which allows us to smooth the discontinuities of the options' payoffs.
Stability analysis of the method is analyzed and performed.
In fact, according to an analysis carried out in the present paper,
the proposed method is unconditionally stable.
Numerical experiments are presented showing that the proposed approaches are extremely accurate and fast.

 \end{abstract}
\maketitle
\section{Introduction}
In the field of numerical methods, the finite element method (FEM) and other mesh-based methods such as finite difference method (FDM) and finite volume method (FVM) are robust and well established. Therefore, they are widely used in financial field due to their applicability to many types of
options, see \cite{Yousuf.Khaliq.Kleefeld, Zvan.Forsyth.Vetzal.thesis, Ballestra.4, Ballestra.Cecere}.
In the FEM, a computational domain is divided into finite elements which are connected together called a mesh.
Although the FEM and the closely related FVM are well-established numerical techniques for computer modeling in engineering and sciences, they are not without shortcoming.
It is well-known that these methods depend strongly on the mesh properties.
However, to compute problems with irregular geometries using these schemes, mesh generation is a far more time consuming and expensive task than solution of the PDEs \cite{WuXueHong.TaoWenQuan}.
\par
To overcome this difficulty associated with FEM and FVM, the boundary element method (BEM) \cite{LucaBallestra.BEM1} appears to be a attractive alternative. In the BEM, only the boundary of domain needs to be discretized \cite{Hosseinzadeh.1,Hosseinzadeh.2,Hosseinzadeh.3,Hosseinzadeh.4}. This reduces the problem dimension by one and thus largely reduces the time in meshing. The BEM still uses elements to implement both interpolation and integration \cite{Hosseinzadeh.1,Hosseinzadeh.2,Hosseinzadeh.3,Hosseinzadeh.4}.
\par
These difficulties can be overcome by the meshless methods (MLM), which have attracted considerable interest over the past decade. MLM also known as meshfree methods. The main advantage of these methods is to approximate the unknown field by a linear combination of shape functions built without having recourse to a mesh of the domain.
In this method, we use a set of nodes
scattered within the domain of the problem as well as sets of
nodes on the boundaries of the domain to represent (not discretize) the domain of the problem and its boundaries \cite{DehghaArezouGhesmati}. These
sets of scattered nodes are called field nodes and they do not form a mesh, meaning it does not require any priori information on
the relation ship between the nodes for the interpolation or approximation of the unknown functions of field variables \cite{G.R.Liu.Y.T.Gu}.
\par
Meshless methods have progressed remarkably in the last decades and some work has been devoted to their classification.
The classification can be done based on different criteria, e.g., formulation procedure, shape function or
the domain representation \cite{SatyaAtluri.ShenShen.mohem}. The formulation procedures are mainly based on the weak from representation (see \cite{Abbasbandy.Ling,Abbasbandy.ANM,Abbasbandy.EABE2,DehghanM.MirzaeiD,DehghanM.MirzaeiD.2}) and the strong form based on the collocation techniques (see \cite{rad6,rad7,Assari.CAM.1,Assari.CAM.2}), although a combination of both approaches is possible \cite{Atluri.1}.
The collocation methods are truly meshless, as the collocation
technique is directly based on a set of nodes without any
background mesh for numerical integration. One limitation of the
collocation methods is less accuracy and lower stability in numerical
implementation.
However, this method is based on point collocation,
and is very sensitive to the choice of collocation points, see \cite{rad.cpc,rad5}.
On the other hands, the weak forms are used to derive a set of algebraic equations through a numerical integration process using a set of quadrature
domain that may be constructed globally or locally in the domain of the problem, such as the element-free
Galerkin method (EFG) \cite{Belytschko.u.Gu}, the reproducing kernel particle method (RKPM) \cite{Liu.Jun.Zhang} and the partition
unity method (PUM) \cite{Melenk.Babuska}.
The above-mentioned methods are all based on a global weak form, being
meshless only in terms of the interpolation of the field or boundary variables, and have to use background cells to integrate
over the problem domain. The requirement of background cells for integration makes these methods being not truly meshless.
In order to alleviate the global integration background cells, the meshless local Petrov-Galerkin method (MLPG) based
on local weak-form and the radial basis functions (RBFs) approximation, was developed by Atluri and his colleagues \cite{S.Atluri.Zhu}. This method also known as local radial point interpolation (LRPI) method. The MLPG method is a truly meshless method, because, a traditional non-overlapping, continuous
mesh is not required, either for the construction of the shape function, or for the integration of the local sub-domain.
The trial and test function spaces can be different or the same. It
offers a lot of flexibility to deal with different boundary value
problems. A wide range of problems has been solved by Atluri
and his coauthors \cite{Atluri.1,Atluri.2}. In this type of MLPG method, the Heaviside step function is employed as a test function.
Particularly, the LRPI meshless
method reduces the problem dimension by one, has shape functions with delta function properties, and expresses the
derivatives of shape functions explicitly and readily. Thus it allows one to easily impose
essential boundary and initial (or final) conditions.
For some works on the meshless local method one can mention
the works of Sladek brothers \cite{shirzadi.LBIE,hosseini.LBIE,sale.LBIE,sladek.1,sladek.2}. The method has now been successfully
extended to a wide rang of problems in engineering. For some
examples of these problems, see \cite{DehghaArezouGhesmati,Dehgahn.LBIE.1} and other references
therein. The interested reader of meshless methods can also
see \cite{Abbasbandy.Ling,Abbasbandy.ANM,Abbasbandy.EABE2}.
\par
Over the last years, the growth of the financial markets has been an ever expanding economical field. In all over the world, the value of the financial
assets traded on the stock markets has reached astronomical amounts.
Trading of financial derivatives such as options is a continuous business going on all over the world. Making sure that the price are correct at every time is of great importance for the traders.\\
Options are financial contracts that gives to the buyer the right, to buy (call option) or to sell (put option) an underlying asset (such as a stock) at a previous agreed price. Called the strike or exercise price on or before a certain time called the maturity. Mast of these options can be grouped into either of the two categories: European options which can only be exercised at one given expiry or maturity date ($t=T$) and American options can additionally be executed at any time prior to their maturity date ($t \leq T$).
The American options give the freedom when to use the option and are often a little bit more expensive than a corresponding European options.\\
The valuation of options lead to mathematical models that are often challenging to solve.
The famous Black-Scholes formula gives an explicit pricing formula for European call and put options on stocks which do not pay dividends, see \cite{Jaillet.Lamberton.Lapeyre}.
The publication in 1973 \cite{Black.Scholes} of the work of Fischer Black and Myron Scholes has been a starting point for the revolution in the option pricing. Their idea was to develop a model based on the assumption that the asset prices follow a geometric Brownian motion.
For European options the Black-Scholes equation results in a boundary value problem of a diffusion equation.
American options pricing is governed by a parabolic partial differential variational inequality (PDVI).
This gives rise to a free boundary problem, see \cite{Hon.1,Hon.2}.
\par
Recently, by using very empirical studies, it has become evident that the assumption of behavior like a log-normal strike diffusion with a constant volatility and a drift in the
standard Black-Scholes model of the underlying asset price is not consistent with that the real market prices of options with various strike prices and
maturities such as volatility smile or skew and heavy tails \cite{Andersen.Andreasen,Toivanen.anm}.
\par
During the last decade, many works have been done to find modifications of classical Black-Scholes model to satisfy these phenomena in financial markets
such as the models with stochastic volatility (SV), the models with jumps (such as Merton and Kou models proposed by Merton and Kou in two different works \cite{R.C.Merton.main} and \cite{S.G.Kou.main}, respectively), their combinations of stochastic volatility and jumps in returns, i.e. stochastic volatility models with jumps (SVJ) introduced by Bates \cite{Bates}, and stochastic volatility
models with jumps in returns and volatility (SVCJ) introduced by Duffie et al. \cite{Duffie}.

In this research, we focus on the SVJ and SVCJ models.
In this work, Merton model is selected to jump term of model.
In Merton's model the asset return follows a standard Wiener process driven by a compound Poisson process with normally distributed jump \cite{Toivanen.anm}.

We have just mentioned such that the mentioned models of course will also lead to
such a volatility smile or skews on short or long term maturity ranges \cite{Ballestra.4}.

The mentioned SVJ and SVCJ models for pricing American options are governed by a parabolic integro-differential variational inequality which can be
formulated as a free boundary problem. In particular, these models are contain differential term and a nonlocal integral term.
Hence, an analytical solution is impossible. Therefore, to solve these problems, we need to have a powerful computational method. To this aim, several numerical methods have been proposed for pricing options under SVJ and SVCJ models (see, e.g., \cite{Ballestra.4,Salmi,Chiarella.Ziogas,Toivanen.Bal}) but
weak form meshless methods have never been used for option pricing of this model, at least to the very best
of our knowledge.
\par
The objective of this paper is to extend the
LRPI based on Wendland's compactly supported radial basis functions (WCS-RBFs) with $C^6$, $C^4$ and $C^2$ smoothness \cite{HOLGER.WENDLAND} to evaluate American options under SVJ and SVCJ models.
Again we do emphasize that, to the best of our knowledge, the local weak form of meshless method has not yet been used in mathematical finance. Therefore, it
appears to be interesting to extend such a numerical technique also to option valuation, which is done in the
presented manuscript.
\par
In addition, in this paper the infinite space domain $\mathbb{R}^+\times \mathbb{R}^+$ is truncated to $[0, S_{max}]\times[0,y_{max}]$ in SVJ and SVCJ models, with the
sufficiently large values $S_{max}$ and $y_{max}$ to avoid an unacceptably large truncation error.
The options' payoffs considered in this paper are non-smooth functions, in particular
 their derivatives are discontinuous at the strike price. Therefore, to reduce as much as possible the losses of accuracy the points of the trial functions are concentrated in a spatial region close to the strike prices. So, we employ the change of variables proposed by Clarke and Parrott \cite{Clarke.Parrott}.
\par
As far as the time discretization is concerned, we use the implicit-explicit (IMEX) time stepping scheme, which is unconditionally
stable and allows us to smooth the discontinuities of the options' payoffs.
Note that in SVJ and SVCJ models, the integral part is a non-local integral, whereas the other parts which are differential operators, are all local.
No doubt, since the integral part is non-local operator, a dense linear system of equations will be obtained by using the $\theta$-weighted discretization scheme.
Therefore, to obtain a sparse linear system of equations, it is better to use a IMEX scheme which is noted for avoiding dense matrices.
So far, and to the best of knowledge, published work existing in the literature which use the IMEX scheme to price the options, include \cite{Ballestra.4,Salmi}.
Such an approach is only first-order
accurate, however a second-order time discretization is obtained by performing a Richardson extrapolation
procedure with halved time step.
Stability analysis of the method is analyzed and performed by the matrix method in the present paper.
\par
Finally, in order to solve the free boundary problem that arises in the case of American options is computed by Richardson extrapolation of the price of Bermudan option. In essence the Richardson extrapolation reduces the free boundary problem and linear complementarity problem to a fixed boundary problem which is much simpler to solve.
\par
Numerical experiments are presented showing that the proposed approach is very efficient from the computational
standpoint. In particular, the prices of both European and American options in SVJ or SVCJ models can be computed
with an error (in both the maximum norm and the root mean square relative
difference) of order $10^{-3}$ in few
tenth of a second. Moreover, the Bermudan approximation reveals to be the most efficient of the algorithms used to deal with the early exercise opportunity.
\par
We remark that the main contribution of this manuscript is to show that the MLPG, which, to the best
of our knowledge, has never been applied to problems in mathematical finance, can yield accurate and fast
approximations of European and American option prices.
\par
Overall , our focus in this paper is more devoted to providing an accurate, computationally fast, stable, convergence and simple technique for pricing
options under SVJ and SVCJ models.
We should note that the key idea of this work is finding a new technique combined using the following numerical tools :
\emph{Spatial change of variables, time discretization of the Black-Scholes operator, Richardson
extrapolation procedure, LRPI discretization, Wendland's compactly supported radial basis functions, Spatial variable transformation , approximate the
price of the American option with the price of a Bermudan option and LU factorization method with partial pivoting.
}
Rigorously speaking, these tools considered separately, are not new, but here due to the fact that an approach that puts all these techniques together has never been proposed in option pricing, therefore the proposed technique is new, accurate and very fast.
\par
The paper is organized as follows: In Section 2 a detailed description of the SVJ and SVCJ models for
American options is provided. Section 3 is devoted to presenting the MLPG approach and the
application of such a numerical technique to the option pricing problems considered is shown in this section. The numerical results obtained are presented and discussed in Section 4 and finally, in Section 5, some
conclusions are drawn.

\section{The option pricing models}\label{model}
\subsection{SVJ or Bates SV model}
For the sake of simplicity, from now we restrict our attention to options of call type, but the case of put
options can be treated in perfect analogy.
\par First we are interested in pricing a American call option of the Bates stochastic volatility model which is the exponential L$\acute{e}$vy processes consisting of a two-dimensional geometric Brownian motion plus a compound Possion jumps with time varying volatility.
\par
Let $(\Omega,\mathcal{F},\mathbf{P})$ be a probability space and also $X_t$ be a continues L$\acute{e}$vy process with a L$\acute{e}$vy measure $\nu$. In the Bates model which is an arbitrage-free market model, the asset price
$S_t$ with time $t\in[0,T]$ and $T$ being the maturity time is than given by
\begin{eqnarray}\nonumber
S_t=S_0e^{X_t},
\end{eqnarray}
where $S_0$ is the asset price at time zero.
\par
Then the risk-neutral dynamics of the asset price $S_t$ and its volatility $Y_t$ are described by the following stochastic differential equations \cite{Ballestra.4,Chiarella.Ziogas,Toivanen.Bal}
\begin{eqnarray}\nonumber
&&\frac{dS_t}{S_t}=\alpha~dt+\sqrt{Y_t}~dW_t^1+dJ_t,\\\nonumber
&&dY_t=\xi(\eta-Y_t)~dt+\theta\sqrt{Y_t}~dW_t^2,
\end{eqnarray}
where $(W_t^1,W_t^2)$ are two Brownian motions with correlation factor $\rho\in [-1,1]$,
$\xi,\eta,\theta \in \mathbb{R}^+$ are mean-reversion rate, long-run mean and the instantaneous volatility of $Y_t$, respectively,
$J_t=\sum_{j=1}^{N_t}R_j$ is a compound Poisson process, where $N_t$ is a Possion process with intensity $\lambda$ and
the set ${R_j}$ is a sequence of independent and identically distributed (i.i.d.) random variables with density $\nu(dx)/\lambda$.
Also $\alpha=r-q-\lambda\kappa$ is the drift rate, where $r$ is the risk-free interest
rate, $q$ is the dividend and $\kappa$ is the expected relative jump size.
Here, we can rewrite the L$\acute{e}$vy measure $\nu(dx)$ as $\lambda f(x)dx$ where $f(x)$ is a weight function. By selecting this weight function the finite activity jump-diffusion model is the log-normal model proposed by Merton \cite{R.C.Merton.main}
\begin{eqnarray}
f(x)=\frac{1}{\sqrt{2\pi}x\delta}\exp(-\frac{(\log x-\gamma)^2}{2\delta^2}),
\end{eqnarray}
Note that $f(x)\geq0$ and
\begin{eqnarray}
\int_{\mathbb{R}^+}f(x)dx=1.
\end{eqnarray}\par
We now consider the pricing of an American call option denoted by $V(s,y,t)$ on the underlying asset $S_t$ with asset price $E$ and maturity $T$.
On can show that $V(s,y,t)$, for $s,y \in [0,+\infty)$ and $t \in [0,T)$, satisfy the following system of free boundary problem
\begin{eqnarray}\label{A.option.1}
&&\frac{\partial}{\partial t}V(s,y,t)+\mathcal{L}V(s,y,t)=0~,~~~~~~~~~~~~~~~~~~~~~0\leq s<B(y,t)~,\\\label{A.option.2}
&&V(s,y,t)=s-E~,~~~~~~~~~~~~~~~~~~~~~~~~~~~~ s>B(y,t)~,\\\label{A.option.3}
&&\lim_{s\rightarrow B(y,t)}\frac{\partial V(s,y,t)}{\partial s}=1~,\\\label{A.option.4}
&&\lim_{s\rightarrow B(y,t)}\frac{\partial V(s,y,t)}{\partial y}=0~,
\end{eqnarray}
where
\begin{eqnarray}\nonumber
&&\mathcal{L}V(s,y,t)=-\mathbf{F}\times\nabla V(s,y,t)+\nabla.(\mathbf{E}\times\nabla V(s,y,t))-(r+\lambda) V(s,y,t)+\lambda\int_{\mathbb{R}^+}V(sx,y,t)f(x)dx,\\\label{bsoperator}
&&\\\nonumber
&&\\\nonumber
&&\mathbf{E}=\frac{1}{2}
\left(
 \begin{array}{cc}
ys^2 & \rho \theta sy\\
\rho\theta sy & \theta^2y
\end{array}
\right)~,\\\nonumber
&&\mathbf{F}=-\left(
 \begin{array}{c}
(r-q-\lambda\kappa)s-ys-\rho\frac{\theta}{2}s\\
\xi(\eta-y)-\frac{\theta^2}{2}-\rho\frac{\theta}{2}y
\end{array}
\right)^T~.\\\nonumber
\end{eqnarray}
Again we should note that $\kappa$ is the expected relative jump size and is computed as $\kappa=\int_{\mathbb{R}^+}(z-1)f(z)dz$.
We have $\kappa=\exp(\gamma+\delta^2/2)-1$ for Merton model.\par
The value of $V$ at maturity is given by
\begin{eqnarray}
V(s,y,T)=\varsigma(s)~,
\end{eqnarray}
 where $\varsigma$ is the so-called option's payoff:
\begin{eqnarray}\label{payoff}
\varsigma(s)=\mathrm{max}(s-E,0)~,
\end{eqnarray}
 which is clearly not differentiable at $s=E$. \par
 The behavior of the value of the American call option on the boundaries is given by
 \begin{eqnarray}\label{bc12}
V(0,y,t)=0~,~~~~~~ \lim_{s\rightarrow +\infty}V(s,y,t)=s-E~~.
\end{eqnarray}
\par
In relations (\ref{A.option.1})-(\ref{A.option.4}), $B(y,t)$ denotes the so-called exercise boundary, which is unknown and is implicitly
 defined by (\ref{A.option.1})-(\ref{bc12}). The above free-boundary partial differential problem does not
  have an exact closed-form solution, and thus some numerical approximation is required. \par
Problem (\ref{A.option.1})-(\ref{bc12}) can be reformulated as a linear complementarity problem:
\begin{eqnarray}\label{A.2}
&&\frac{\partial}{\partial t}V(s,y,t)+\mathcal{L}V(s,y,t)\geq0~,\\\label{A.222}
&&V(s,y,t)-\varsigma(s)\geq0~,\\
&&\left(\frac{\partial}{\partial t}V(s,y,t)+\mathcal{L}V(s,y,t)\right) \cdot \left(V(s,y,t)-\varsigma(s)\right)=0~,\label{A.2final}
\end{eqnarray}
which holds for $s,y \in (0,+\infty)$ and $t \in [0,T)$, with final condition:
\begin{eqnarray}\label{ini2}
V(s,y,T)=\varsigma(s)~,
\end{eqnarray}
and boundary conditions:
\begin{eqnarray}\label{bc2}
V(0,y,t)=0~,~~~~~~ \lim_{s\rightarrow +\infty}V(s,y,t)=s-E~~.
\end{eqnarray}
\par
\subsection{SVCJ model}
In the Bates models, Poisson jumps are only added to the risk-neutral dynamic of the asset price $S_t$. In contrast, in the SVCJ models,
jumps are appeared in the volatility, too. Then we have \cite{Salmi}
\begin{eqnarray}\label{SVCJ.1}
&&\frac{dS_t}{S_t}=\alpha_s~dt+\sqrt{Y_t}~dW_t^1+dJ^1_t,\\\label{SVCJ.2}
&&dY_t=\xi(\eta-Y_t)~dt+\theta\sqrt{Y_t}~dW_t^2+dJ^2_t,
\end{eqnarray}
where $\alpha_s=r-q-\lambda\kappa_s$ such that $\kappa_s=(1-\nu\rho_j)^{-1}\exp(\gamma+\delta^2/2)-1$ and $\rho_j$
defines the correlation between jumps in returns and variance. The two-dimensional jump
process $(J^1,J^2)$ is a $\mathbb{R}\times \mathbb{R}^+$-valued compound Poisson process with intensity $\lambda$ \cite{Salmi}.
The distribution of the jump size in variance is assumed to be exponential with mean $\nu$.
Conditional on a jump of size $z^v$ in the variance process, $J^1+1$ has a log-normal distribution
$f(z^s,z^v)$ with the mean in log $z^s$ being $\gamma+\rho_jz^v$ \cite{Salmi}. This gives a bivariate probability density
function defined by \cite{Salmi}
$$f(z^s,z^v)=\frac{1}{\sqrt{2\pi}z^s\delta\nu}\exp\Bigg(-\frac{z^v}{\nu}-\frac{(\log z^s -\gamma-\rho_jz^v)^2}{2\delta^2}\Bigg).$$
Let $V(s,y,t)$ denote the price of a American derivative on an underlying asset described by model \ref{SVCJ.1} and \ref{SVCJ.2}. As in \cite{Salmi}, it can be
shown that $V(s,y,t)$ is governed by the partial integro-differential equation
\begin{eqnarray}\label{A.option.1e}
&&\frac{\partial}{\partial t}V(s,y,t)+\mathcal{L}V(s,y,t)=0~,~~~~~~~~~~~~~~~~~~~~~0\leq s<B(y,t)~,\\\label{A.option.2e}
&&V(s,y,t)=s-E~,~~~~~~~~~~~~~~~~~~~~~~~~~~~~ s>B(y,t)~,\\\label{A.option.3e}
&&\lim_{s\rightarrow B(y,t)}\frac{\partial V(s,y,t)}{\partial s}=1~,\\\label{A.option.4e}
&&\lim_{s\rightarrow B(y,t)}\frac{\partial V(s,y,t)}{\partial y}=0~,
\end{eqnarray}
where
\begin{eqnarray}\nonumber
&&\mathcal{L}V(s,y,t)=-\mathbf{F}\times\nabla V(s,y,t)+\nabla.(\mathbf{E}\times\nabla V(s,y,t))-(r+\lambda) V(s,y,t)\\\nonumber
&&+\lambda\int_{\mathbb{R}^+}\int_{\mathbb{R}^+}V(sz^s,y+z^v,t)f(z^s,z^v)dz^v~dz^s,\\\label{bsoperatorSVCJ}
&&\\\nonumber
&&\\\nonumber
&&\mathbf{E}=\frac{1}{2}
\left(
 \begin{array}{cc}
ys^2 & \rho \theta sy\\
\rho\theta sy & \theta^2y
\end{array}
\right)~,\\\nonumber
&&\mathbf{F}=-\left(
 \begin{array}{c}
(r-q-\lambda\kappa)s-ys-\rho\frac{\theta}{2}s\\
\xi(\eta-y)-\frac{\theta^2}{2}-\rho\frac{\theta}{2}y
\end{array}
\right)^T~,\\\nonumber
\end{eqnarray}
the final and boundary conditions for this model are described in relations (\ref{ini2}) and (\ref{bc2}).
\section{Methodology}\label{Methodology}
In this work, the price of American option is computed by Richardson extrapolation of the price of Bermudan option. In essence the Richardson extrapolation reduces the free boundary problem and linear complementarity problem to a fixed boundary problem which is much simpler to solve. Thus, instead of describing the aforementioned linear complementarity problem or penalty method, we directly focus our attention onto the partial integro-differential equation satisfied by the price of a Bermudan option which is faster and more accurate than other methods.

For the sake of simplicity exposition, we restrict our attention to option of the Bates stochastic volatility model, but the case of options under SVCJ model can be treated in perfect analogy.

Let us consider in the interval $[0,T]$, $M+1$ equally spaced time levels $t_0=0,t_1,t_2,...,t_M=T$. Let $V_M(s,y,t)$ denote the price of a Bermudan option with maturity $T$ and strike price $E$.
The Bermudan option is an option
 that can be exercised not on the whole time interval $[0,T]$, but only at the dates $t_0$, $t_1$, $\ldots$, $t_M$.
That is we consider the problems

\begin{eqnarray}
\begin{cases}
\frac{\partial}{\partial t}V_M(s,y,t)+\mathcal{L}V_M(s,y,t) = 0~, \label{A.22} \\
V_M(0,y,t)=0~,~~~~~~ \lim_{s\rightarrow +\infty}V_M(s,y,t)=s-E~.
\end{cases}
\end{eqnarray}

which hold in the time intervals $(t_{0},t_{1})$, $(t_{1},t_{2})$, $\ldots$, $(t_{M-1},t_{M})$. By doing that also the relation (\ref{A.2final}) is automatically satisfied in every time interval $(t_{k},t_{k+1})$,
   $k=0,1,\ldots,M-1$. Moreover, the relation
  (\ref{A.222}) is enforced only at times $t_0$, $t_1$, $\ldots$, $t_{M-1}$, by setting

\begin{eqnarray}\label{A.22.Ame}
V_M(s,y,t_k) = \max(\lim_{t\rightarrow t_k^+}V_M(s,y,t),\varsigma(s))~, ~~~~~k=0,1,\ldots,M-1.~
\end{eqnarray}

Note that the function $V_M(\cdot,\cdot,t_{k})$ computed according to (\ref{A.22.Ame}) is used as the final condition for the problem (\ref{A.22})
  that holds in the time interval  $(t_{k-1},t_{k})$, $k=1,2,\ldots,M-1$. Instead, the final condition
   for the problem (\ref{A.22}) that holds in the time interval $(t_{M-1},t_{M})$,
   according to the relation (\ref{ini2}), is prescribed as follows:

\begin{eqnarray}\label{A.22.Final}
V_M(s,y,t_M) = \varsigma(s)~.
\end{eqnarray}

That is, in summary, problems (\ref{A.22}) are recursively solved for $k=M-1,M-2,\ldots,0$,
 starting from the condition (\ref{A.22.Final}), and at each time $t_{M-1}$, $t_{M-2}$, $\ldots$, $t_0$
  the American constraint (\ref{A.22.Ame}) is imposed. \par

The Bermudan option price $V_M(s,y,t)$ tends to become a fair approximation of the American option price $V(s,y,t)$ as
the number of exercise dates $M$ increases. In this work the accuracy of $V_M(s,y,t)$ is enhanced by Richardson extrapolation which is second-order accurate in time.
\\\\\\
To evaluate the option, the meshless local Petrov-Galerkin (MLPG) method is used in this work.
This method is based on local weak
forms over intersecting sub-domains, which are extracted over the
local sub-domains using divergence theorem and a Heaviside test
function. At first we discuss a time-stepping method for the
time derivative.
\subsection{Time discretization}
First of all, we discretize the operator (\ref{bsoperator}) in time. For this propose, we can apply the Laplace transform or use a time-stepping
approximation. Algorithms for the numerical inversion of a Laplace transform lead to a reduction in accuracy. Then, we employ a time-stepping method to
overcome the time derivatives in this operator.

Let $V^k(s,y)$ denote a function approximating $V_M(s,y,t_k), k=0,1,...,M-1$. Note that the subscript $M$ has been removed
from $V^k(s,y)$ to keep the notation simple. According to (\ref{A.22.Final}), we set $V^M(s,y)=\varsigma(s)$.
Let us consider the following implicit-explicit (IMEX) time stepping scheme:

\begin{eqnarray}
\mathcal{L}V^{k}(s,y)= -\mathbf{F}\times\nabla V^k(s,y)+\nabla.(\mathbf{E}\times\nabla V^k(s,y))-(r+\lambda) V^k(s,y)+\lambda\int_{\mathbb{R}^+}V^{k+1}(sx,y)f(x)dx,\label{bsoperator2}
\end{eqnarray}
and also we use
\begin{eqnarray}\label{partialt}
\frac{\partial}{\partial t}V(s,y,t)\simeq\frac{V^{k+1}(s,y)-V^{k}(s,y)}{\Delta t}+\mathcal{O}(\Delta t),
\end{eqnarray}
using relations (\ref{bsoperator2}) and (\ref{partialt}), we define the following operator
\begin{eqnarray}\nonumber
&&\mathcal{\widetilde{L}}V^{k}(s,y)\simeq \frac{V^{k+1}(s,y)-V^{k}(s,y)}{\Delta t}+\mathcal{L}V^{k}(s,y)\\\nonumber
&&=\frac{1}{\Delta t}V^{k+1}(s,y)-\mathbf{F}\times\nabla V^k(s,y)+\nabla.(\mathbf{E}\times\nabla V^k(s,y))
-(r+\lambda+\frac{1}{\Delta t}) V^k(s,y)+\lambda\int_{\mathbb{R}^+}V^{k+1}(sx,y)f(x)dx,\label{bsoperator22}
\end{eqnarray}

Therefore, the American option problems are rewritten as follows:
\begin{eqnarray}\label{baghorban}
\begin{cases}
\mathcal{\widetilde{L}}V^{k}(s,y) = 0~,  \\
V^k(0,y)=0~,~~~~~~ \lim_{s\rightarrow +\infty}V^k(s,y)=s-E~~.
\end{cases}
\end{eqnarray}
and also, the relations (\ref{A.22.Ame}) and (\ref{A.22.Final}) are rewritten
as follows:
\begin{eqnarray}\label{eq.final.amer}
&&V^{k}(s,y) = \max(\lim_{t\rightarrow t_{k+1}^+}V^{k}(s,y),\varsigma(s))~, ~~~~~k=M-1,M-2,\ldots,1,0~\\\nonumber
&&V^{M}(s,y) = \varsigma(s)~.
\end{eqnarray}
\textbf{Remark~1:}
Note that in relation (\ref{bsoperator2}), the integral part is a non-local integral, whereas the other parts which are differential operators, are all local.
No doubt, since the integral part is non-local operator, a dense linear system of equations will be obtained by using the $\theta$-weighted discretization scheme.
Therefore, to obtain a sparse linear system of equations, it is better to use a IMEX scheme which is noted for avoiding dense matrices.
So far, and to the best of knowledge, published work existing in the literature which use the IMEX scheme to price the options, include \cite{Salmi}.
Therefore, in this work, we use the IMEX scheme which is only first-order accurate in time. Then,
the obtained approximation, which is only first-order accurate, is
improved by Richardson extrapolation. In particular, we manage to
obtain second-order accuracy by extrapolation of two solutions
computed using $M$ and $2M$ time steps.
In the following, for the sake of brevity, we will restrict our
attention to first stage of the Richardson extrapolation
procedure, where $M$ time steps are employed, and the fact that
the partial integro-differential problems considered are also solved with
$2M$ time steps will be understood.
\subsection{Spatial variable transformation}
\begin{figure}
\center
\includegraphics[width=16cm,height=10cm]{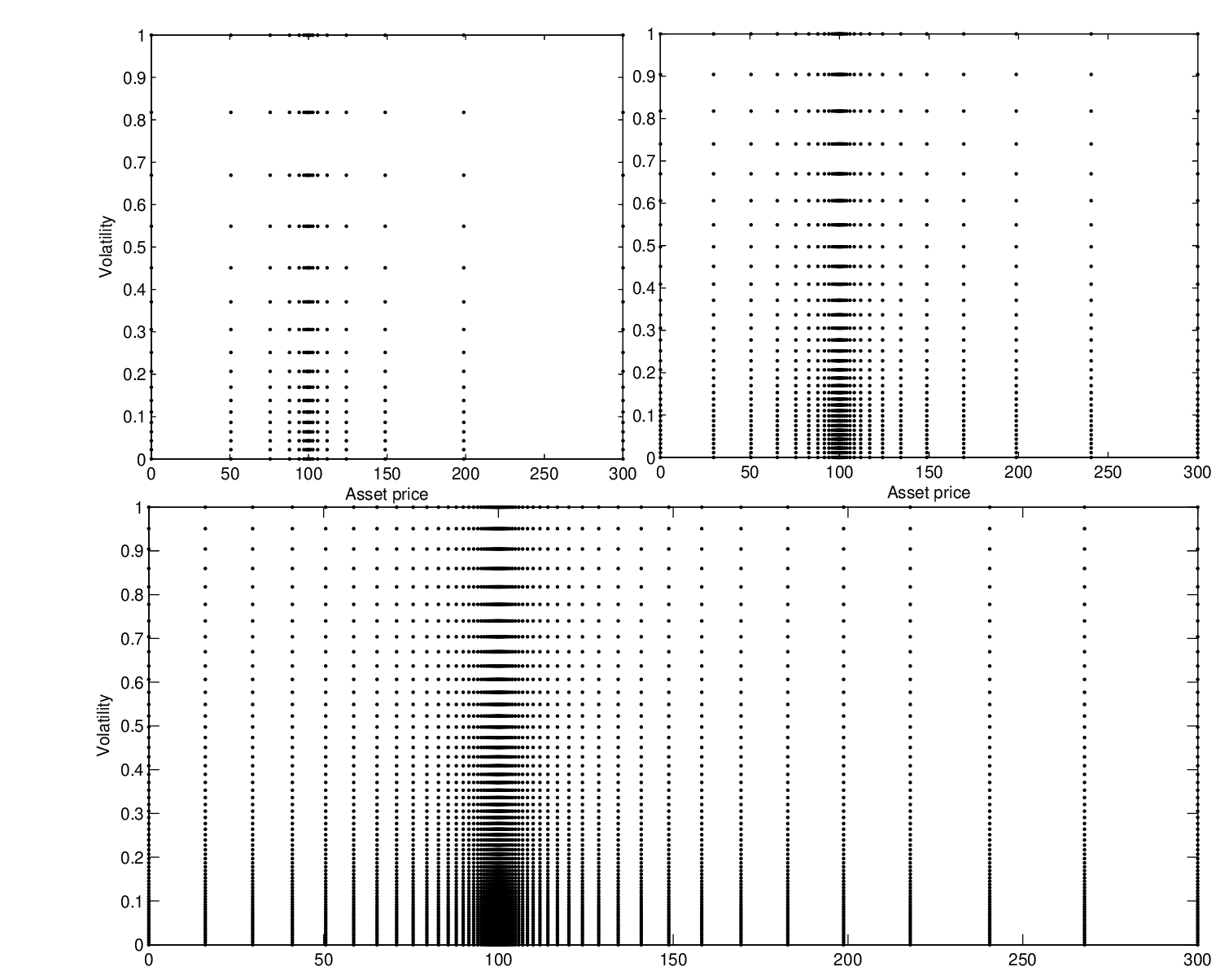}
\caption{$\xi_s=1$, $\xi_y=10$, $N_x,N_z=16,32,64$, respectively. (Note that $N_x$ and $N_z$ are introduced in relation \ref{nodes}.)} \label{nodesfig}
\end{figure}

It is well-known that from the mathematical point of view, the Bates stochastic volatility model typically leads to a partial integro-differential equation
that is defined in the unbounded spatial domain $\mathbb{R}^+\times\mathbb{R}^+$. But due to the fact that it requires large memory storage, we replace the domain
with the finite domain $\Omega=[0,S_{max}]\times[0,y_{max}]$ of the asset price and the volatility, where $S_{max}$ and $y_{max}$ are chosen sufficiently
large to avoid an unacceptably large
truncation error. However, in \cite{WilmottHowison} shown that
upper bound of the asset price is three or four times of the
strike price, so we can set $S_{max}=4E$. The options' payoffs
considered in this paper are non-smooth functions, in particular
their derivatives are discontinuous at the strike price.
Therefore, to reduce the losses of accuracy the points of the
trial functions are concentrated in a spatial region close to
$s=E$.
In contrast, along the $y$-direction, we want to have a mesh which is finer in a neighborhood of $y = y_0$, where the possible realizations of the
variance process are more likely to occur \cite{Ballestra.4}.
So, we employ the following change of variables:
\begin{eqnarray}\label{changeofv}
&&x(s)=\frac{\sinh^{-1}(\xi_s(s-E))+\sinh^{-1}(\xi_s E)}{\sinh^{-1}(\xi_s(S_{max}-E))+\sinh^{-1}(\xi_s E)},\\\label{changeofv2}
&&z(y)=\frac{\sinh^{-1}(\xi_y(y-y_0))+\sinh^{-1}(\xi_y y_0)}{\sinh^{-1}(\xi_y(y_{max}-y_0))+\sinh^{-1}(\xi_y y_0)},
\end{eqnarray}
or
\begin{eqnarray}\label{strans}
&&s(x)=\frac{1}{\xi_s}\sinh\Bigg(x\sinh^{-1}(\xi_s(S_{max}-E))-(1-x)\sinh^{-1}(\xi_s E)\Bigg)+E,\\\nonumber
&&y(z)=\frac{1}{\xi_y}\sinh\Bigg(z\sinh^{-1}(\xi_y(y_{max}-y_0))-(1-z)\sinh^{-1}(\xi_y y_0)\Bigg)+y_0,\\\nonumber
\end{eqnarray}
where $\xi_s$ and $\xi_y$ are the suitable constant parameters. Using these parameters, we can control the amount of the distribution of
nodes in the $s$ and $y$-directions near $s=E$ and $y=y_0$, see Figure \ref{nodesfig}.
\par
Note that the relations (\ref{changeofv}) and (\ref{changeofv2}) maps the $[0, S_{max}]\times[0,y_{max}]$ to the $[0, 1]\times[0,1]$.
We define
\begin{eqnarray}\\\nonumber
&&U(x,z,t)  =  V(s(x),y(z),t),\\\nonumber
&&\widetilde{\mathcal{L}}U^{k}(x,z)  =  \frac{1}{\Delta t}U^{k+1}(x,z)-(\widetilde{\mathbf{F}}*\mathbf{P})\times\nabla U^k(x,z)+\mathbf{P}\times[\nabla*([\widetilde{\mathbf{E}}*\mathbf{P}^T]^T\times\nabla U^k(x,z))]\\\nonumber
&&~~~~~~~~~~~~~~~-(r+\lambda+\frac{1}{\Delta t}) U^k(x,z)\\\label{Lhatop}
&&~~~~~~~~~~~~~~~+\lambda\int_{0}^{1}U^{k+1}(\widehat{r},z)f(\frac{r(\widehat{r})}{s(x)})\frac{1}{s(x)}r'(\widehat{r})d\widehat{r}
+\lambda\int_{S_{max}}^{\infty}(r-E)f(\frac{r}{s(x)})\frac{1}{s(x)}dr,
\end{eqnarray}
where the symbol $*$ means component-wise multiplication and also
\begin{eqnarray}\label{bsoperatorr}
&&\widetilde{\mathbf{E}}=\frac{1}{2}
\left(
 \begin{array}{cc}
y(z)s^2(x) & \rho \theta s(x)y(z)\\
\rho\theta s(x)y(z) & \theta^2y(z)
\end{array}
\right)~,\\\nonumber
&&\widetilde{\mathbf{F}}=-\left(
 \begin{array}{c}
(r-q-\lambda\kappa)s(x)-y(z)s(x)-\rho\frac{\theta}{2}s(x)\\
\xi(\eta-y(z))-\frac{\theta^2}{2}-\rho\frac{\theta}{2}y(z)
\end{array}
\right)^T~,\\\nonumber
&&\mathbf{P}=\left(
 \begin{array}{c}
\frac{1}{s'(x)}\\
\frac{1}{y'(z)}
\end{array}
\right)^T~,\\\nonumber
&&r(\widehat{r})=\frac{1}{\xi_s}\sinh\Bigg(\widehat{r}\sinh^{-1}(\xi_s(S_{max}-E))-(1-\widehat{r})\sinh^{-1}(\xi_s E)\Bigg)+E,\\\nonumber
\end{eqnarray}
Using the change of variable (\ref{changeofv}), the relations (\ref{baghorban}) are rewritten as follows:
\begin{eqnarray}\label{eq.main}
\begin{cases}
\widetilde{\mathcal{L}}U^{k}(x,z) = 0~,  \\
U^k(0,z)=0~,~~~~~~ U^k(1,z)=S_{max}-E~~.
\end{cases}
\end{eqnarray}
and also, the relations (\ref{eq.final.amer}) are rewritten
as follows:
\begin{eqnarray}\label{eq.final.amer.tran}
&&U^{k}(x,z) = \max(\lim_{t\rightarrow t_{k+1}^+}U^{k}(x,z),\widetilde{\varsigma}(x))~, ~~~~~k=0,1,\ldots,M-1,~\\\nonumber
&&U^{M}(x,z) = \widetilde{\varsigma}(x)~,
\end{eqnarray}
where
\begin{eqnarray}\label{payoff.tran}
\widetilde{\varsigma}(x)=\mathrm{max}(s(x)-E,0)~.
\end{eqnarray}
\subsection{The local weak form}
\begin{figure}
\center
\includegraphics[width=10cm,height=6cm]{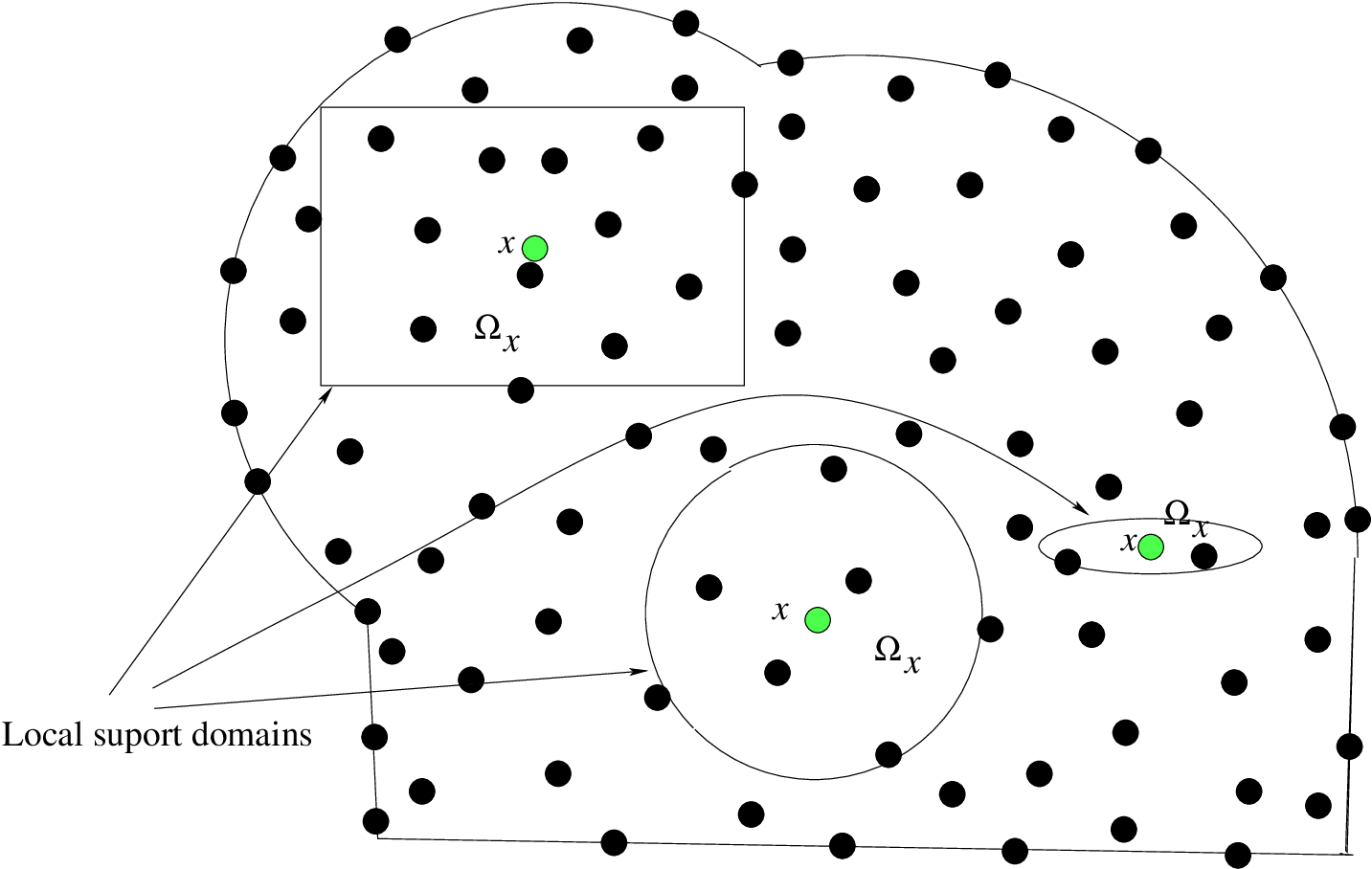}
\caption{Support sub-domains of different points in the problem domain.} \label{fig.prob.domain}
\end{figure}

In this section, we use the local weak form instead of the global
weak form. The local weak form meshless method was firstly proposed by Atluri et
al. \cite{SatyaAtluri.ShenShen.mohem}, in their meshless local
Petrov-Galerkin method. The MLPG method constructs the weak form
over local sub-domains such as $\Omega_s$, which is a small region
taken for each node in the global domain $\Omega=[0,1]\times [0,1]$. The local
sub-domains overlap each other and cover the whole global domain
$\Omega$. This local sub-domains could be any simple geometry like a circle, square, as shown in
Figure \ref{fig.prob.domain}.
For simplicity, we assume the local sub-domains have circular shape.
Therefore the local weak form of the approximate equation (\ref{eq.main})
for $\mathbf{x} \in \Omega_{s}^{i}$, where $\mathbf{x}=(x,z)$, can be written as
\begin{eqnarray}\label{eq.main.2}
<\widetilde{\mathcal{L}}U^{k},u^{\star}> = 0~,
\end{eqnarray}
In equation (\ref{eq.main.2}), $u^{\star}$ is the Heaviside step function
\begin{eqnarray}\nonumber
u^{\star}(\mathbf{x})=
\begin{cases}
1 & \mathbf{x}\in \Omega_s^i,\\
0 & o.w.
\end{cases}
\end{eqnarray}
as the test function in each local domain. Also we define the inner product $<.,.>$ on interior domain and $\{.,.\}$ on boundary as
\begin{eqnarray}\label{eq.main.3}
&&<\widetilde{\mathcal{L}}U^{k},u^{\star}>= \int_{\Omega_s^i}\widetilde{\mathcal{L}}U^{k}(\mathbf{x})u^{\star}(\mathbf{x})d\Omega~,\\\nonumber
&&\{\widetilde{\mathcal{L}}U^{k},u^{\star}\}= \int_{\partial\Omega_s^i}\widetilde{\mathcal{L}}U^{k}(\mathbf{x})u^{\star}(\mathbf{x})d\Gamma~,
\end{eqnarray}
 which $\Omega_s^i$ is the local domain associated with the point
$i$, i.e. it is a circle centered at $\mathbf{x}$ of radius $r_Q$. Let $\partial\Omega_s^i$ denote the boundary of sub-domain $\Omega_s^i$.\par
Eq. (\ref{eq.main.3}) with the substitution relation (\ref{Lhatop}) constitute the following relation
\begin{eqnarray}\nonumber
&&<(\widetilde{\mathbf{F}}*\mathbf{P})\times\nabla U^k,u^\star>-
<\mathbf{P}\times[\nabla*([\widetilde{\mathbf{E}}*\mathbf{P}^T]^T\times\nabla U^k(x,z))],u^\star>+(r+\lambda+\frac{1}{\Delta t})<U^k,u^\star>\\\label{hama}
&&=\frac{1}{\Delta t}<U^{k+1},u^\star>+\lambda<\mathcal{W},u^\star>+\lambda<\Pi,u^\star>
\end{eqnarray}
where
\begin{eqnarray}\nonumber
&&\mathcal{W}(z)=\int_{0}^{1}U^{k+1}(\widehat{r},z)f(\frac{r(\widehat{r})}{s(x)})\frac{1}{s(x)}r'(\widehat{r})d\widehat{r},\\\nonumber
&&\Pi=\int_{S_{max}}^{\infty}(r-E)f(\frac{r}{s(x)})\frac{1}{s(x)}dr.
\end{eqnarray}
\par
An alternative for $<\mathbf{P}\times[\nabla*([\widetilde{\mathbf{E}}*\mathbf{P}^T]^T\times\nabla U^k(x,z))],u^\star>$ related to relation (\ref{hama}) can be mentioned as
\begin{eqnarray}\nonumber
&&<\mathbf{P}\times[\nabla*([\widetilde{\mathbf{E}}*\mathbf{P}^T]^T\times\nabla U^k(x,z))],u^\star>=
<[\mathbf{P}*[\nabla*(\mathbf{P}\times\widetilde{\mathbf{E}}_1)+\widetilde{\mathbf{E}}_2]]\times\nabla U^k,u^\star>\\\nonumber
&&~~~~~~~~~~~~~~~~~~~~~~~~~~~~~~~~~~~~~~~+<[\texttt{Diag}(\widetilde{\mathbf{E}})*\mathbf{P}*\mathbf{P}]\times\Delta. U^k,u^\star>+\rho\theta<\frac{s(x)y(z)}{s'(x)y'(z)}\frac{\partial^2}{\partial x\partial z}U^{k},u^\star>,
\end{eqnarray}
where
\begin{eqnarray}\label{bsoperatorrr}
&&\widetilde{\mathbf{E}}_1=\frac{1}{2}
\left(
 \begin{array}{cc}
y(z)s^2(x) & 0\\
0 & \theta^2y(z)
\end{array}
\right)~,\\\nonumber
&&\widetilde{\mathbf{E}}_2=\frac{1}{2}\rho\theta\left(
 \begin{array}{c}
s(x)\\
y(z)
\end{array}
\right)^T~,\\\nonumber
&&\Delta. U^k=\left(
 \begin{array}{c}
\frac{\partial^2}{\partial x^2}U^{k}\\
\frac{\partial^2}{\partial z^2}U^{k}
\end{array}
\right)^T~.
\end{eqnarray}
What will be used in here for simplifying the system (\ref{hama}) is the divergence theorem as follows
\begin{eqnarray}\nonumber
&&<(\widetilde{\mathbf{F}}*\mathbf{P})\times\nabla U^k,u^\star>=-<[\nabla.(\widetilde{\mathbf{F}}*\mathbf{P})]~ U^k,u^\star>
+\{[(\widetilde{\mathbf{F}}*\mathbf{P}).\mathbf{\nu}]U^k,u^\star\},\\\nonumber
&&<[\mathbf{P}*[\nabla*(\mathbf{P}\times\widetilde{\mathbf{E}}_1)+\widetilde{\mathbf{E}}_2]]\times\nabla U^k,u^\star>=
-<\nabla.[\mathbf{P}*[\nabla*(\mathbf{P}\times\widetilde{\mathbf{E}}_1)+\widetilde{\mathbf{E}}_2]]~ U^k,u^\star>\\\nonumber
&&~~~~~~~~~~~~+\{[\mathbf{P}*[\nabla*(\mathbf{P}\times\widetilde{\mathbf{E}}_1)+\widetilde{\mathbf{E}}_2]].\mathbf{\nu} ~U^k,u^\star\},\\\nonumber
&&<[\texttt{Diag}(\widetilde{\mathbf{E}})*\mathbf{P}*\mathbf{P}]\times\Delta. U^k,u^\star>
=<\Delta.(\texttt{Diag}(\widetilde{\mathbf{E}})*\mathbf{P}*\mathbf{P})~U^k,u^\star>
-\{[[\nabla*[\texttt{Diag}(\widetilde{\mathbf{E}})*\mathbf{P}*\mathbf{P}]].\mathbf{\nu}]~U^k,u^\star\}\\\nonumber
&&~~~~~~~~~~~~+\{[[\texttt{Diag}(\widetilde{\mathbf{E}})*\mathbf{P}*\mathbf{P}]* \nabla U^k].\mathbf{\nu},u^\star\},\\\nonumber
&&<\frac{s(x)y(z)}{s'(x)y'(z)}\frac{\partial^2}{\partial x\partial z}U^{k},u^\star>=
<(\frac{s(x)}{s'(x)})'(\frac{y(z)}{y'(z)})'U^{k},u^\star>-\{(\frac{s(x)}{s'(x)})(\frac{y(z)}{y'(z)})'U^{k}\nu^1,u^\star\}\\\label{diverg}
&&~~~~~~~~~~~~+\{\frac{s(x)y(z)}{s'(x)y'(z)}\frac{\partial}{\partial x}U^{k}\nu^2,u^\star\},
\end{eqnarray}
where $\mathbf{\nu}$ is unit outward normal vector on the boundary of the
domain of the problem.
Substituting relations (\ref{diverg}) in (\ref{hama}), we obtain
\begin{eqnarray}\nonumber
&&-<[\nabla.(\widetilde{\mathbf{F}}*\mathbf{P})]~ U^k,u^\star>
+<\nabla.[\mathbf{P}*[\nabla*(\mathbf{P}\times\widetilde{\mathbf{E}}_1)+\widetilde{\mathbf{E}}_2]]~ U^k,u^\star>\\\nonumber
&&-<\Delta.(\texttt{Diag}(\widetilde{\mathbf{E}})*\mathbf{P}*\mathbf{P})~U^k,u^\star>
-<(\frac{s(x)}{s'(x)})'(\frac{y(z)}{y'(z)})'U^{k},u^\star>\\\nonumber
&&+(r+\lambda+\frac{1}{\Delta t})<U^k,u^\star>
+\{[(\widetilde{\mathbf{F}}*\mathbf{P}).\mathbf{\nu}]U^k,u^\star\}
-\{[\mathbf{P}*[\nabla*(\mathbf{P}\times\widetilde{\mathbf{E}}_1)+\widetilde{\mathbf{E}}_2]].\mathbf{\nu} ~U^k,u^\star\}\\\nonumber
&&+\{[[\nabla*[\texttt{Diag}(\widetilde{\mathbf{E}})*\mathbf{P}*\mathbf{P}]].\mathbf{\nu}]~U^k,u^\star\}
-\{[[\texttt{Diag}(\widetilde{\mathbf{E}})*\mathbf{P}*\mathbf{P}]* \nabla U^k].\mathbf{\nu},u^\star\}\\\nonumber
&&+\{(\frac{s(x)}{s'(x)})(\frac{y(z)}{y'(z)})'U^{k}\nu^1,u^\star\}-\{\frac{s(x)y(z)}{s'(x)y'(z)}\frac{\partial}{\partial x}U^{k}\nu^2,u^\star\}\\\label{finalweak}
&&=\frac{1}{\Delta t}<U^{k+1},u^\star>+\lambda<\mathcal{W},u^\star>+\lambda<\Pi,u^\star>
\end{eqnarray}
It is important to observe that in relation (\ref{finalweak}) exist unknown functions, we should approximate these functions. To this aim the local integral equations (\ref{finalweak}) are transformed in to a system of algebraic equations with real unknown quantities at nodes used for spatial approximation, as described in
the next subsection.

\subsection{Spatial approximation}\label{MLSsection}
Rather than using traditional non-overlapping, contiguous meshes
to make the interpolation scheme, the MLPG method uses a local
interpolation or approximation to represent the trial or test
functions with the values (or the fictitious values) of the
unknown variable at some randomly located nodes. We will find a
number of local interpolation schemes for this purpose. The radial point interpolation method is certainly one of them. The LRPI scheme
is utilized in this paper. In this section, the fundamental idea
of the LRPI is reviewed.

Consider a subdomain $\Omega_\mathbf{x}$ of
$\Omega=[0,1]\times [0,1]$ in the neighborhood of a point $\mathbf{x}$ for the definition of
the LRPI approximation of the trial function around $\mathbf{x}$.

 According to the local point interpolation \cite{G.R.Liu.Y.T.Gu}, the value of point interpolation approximation of $U^k(\mathbf{x})$ at any (given) point $\mathbf{x} \in \Omega$
 is approximated by interpolation at $n$ nodes $\mathbf{x}_1,\mathbf{x}_2$,
  $\ldots$, $\mathbf{x}_n$ (centers) laying
 in a convenient neighborhood of $\mathbf{x}$ i.e. $\Omega_{\mathbf{x}}$.
 The domain in which these nodes are chosen, whose
  shape may depend on the point $\mathbf{x}$,
 is usually referred to as local support domain.
 Various different local point interpolation approaches can be obtained depending on the functions used to interpolate
   $U^k(\mathbf{x})$. In this paper we focus our attention onto the so-called local radial point interpolation
    method (LRPI), which employs a combination of polynomials
     and radial basis functions.
     \par
To
approximate the distribution of function $U^{k}(\mathbf{x})$ in $\Omega_\mathbf{x}$,
over a number of randomly located nodes $\{\mathbf{x}_i\}, ~~i=1,2,...,n$,
the radial point interpolation approximation $\widetilde{U}^k(\mathbf{x})$ of
$U^k(\mathbf{x})$ for each $\mathbf{x} \in \Omega_\mathbf{x}$, can be defined by
\begin{eqnarray}\label{RPI}
\widetilde{U}^k(\mathbf{x})=\sum_{i=1}^{n}R_i(\mathbf{x})a_i^k+\sum_{j=1}^{m}P_j(\mathbf{x})b_j^k~,
\end{eqnarray}

 where $P_1$, $P_2$, $\ldots$, $P_m$ denote the first $m$ monomials in ascending order and $R_1$,
  $R_2$, $\ldots$, $R_n$ are $n$ radial functions centered at $\mathbf{x}_1$, $\mathbf{x}_2$, $\ldots$, $\mathbf{x}_n$, respectively. Moreover $a_1^k$, $a_2^k$,
     $\ldots$, $a_n^k$, $b_1^k$, $b_2^k$, $\ldots$, $b_m^k$ are $n+m$ real coefficients  that have
      to be determined. \par

As far as the radial basis functions $R_1$,
  $R_1$, $\ldots$, $R_n$ are concerned, several choices are possible (see, for example,
  \cite{buhmann.book}).
  In this work we decide to use the Wendland's compactly supported radial basis functions (WCS-RBFs) with $C^6$, $C^4$ and $C^2$ smoothness \cite{HOLGER.WENDLAND}, as they do not involve any free shape parameter (which is not  straightforward to choose, see \cite{Cheng.Golberg.Kansa.Zammito, Carlson.Foley, G.E.Fasshauer.J.G.Zhang, Ballestra.2, Ballestra.3}).
  WCS-RBFs with $C^6$, $C^4$ and $C^2$ smoothness degrees are as follows, respectively:
\begin{eqnarray}\nonumber
&&R_i(s)=(1-r_i)^4_{+}(1+4r_i),~~~~~i=1,2,\ldots,n~,\\\nonumber
&&R_i(s)=(1-r_i)^6_{+}(3+18r_i+35r_i^2),~~~~~i=1,2,\ldots,n~,\\\nonumber
&&R_i(s)=(1-r_i)^8_{+}(1+8r_i+25r_i^2+32r_i^3),~~~~~i=1,2,\ldots,n~,
\end{eqnarray}
where $r_i=\|\mathbf{x}-\mathbf{x}_i\|/r^i_w$ is the distance from node $\mathbf{x}_i$ to $\mathbf{x}$, while $r^i_w$ is the size of support for the radial function $R_i(\mathbf{x})$. In this study, for simplicity, we set $r^i_w=r_w$ for all $i$.
Also, $(1-r_i)^l_{+}$ is $(1-r_i)^l$ for $0\leq r_i<1$ and zero otherwise.\par
Note that the monomials $P_1$, $P_2$, $\ldots$, $P_m$  are not always employed (if $b_i^k=0$, $i=1,2,\ldots,m$, pure RBF approximation is obtained).
 In the present work, both the constant and the linear monomials are used to augment the RBFs (i.e. we set $m=4$).
 \par By requiring that the function $\widetilde{U}^k$ interpolate $U$ at $\mathbf{x}_1$, $\mathbf{x}_2$,
  $\ldots$, $\mathbf{x}_n$, we obtain
  a set of $n$ equations in the $n+m$ unknown coefficients $a_1^k$, $a_2^k$, $\ldots$, $a_n^k$, $b_1^k$,
  $b_2^k$, $\ldots$, $b_m^k$:

  \begin{eqnarray}\label{RPI2}
\sum_{i=1}^{n}R_i(\mathbf{x}_p)a_i^k+\sum_{j=1}^{m}P_j(\mathbf{x}_p)b_j^k=\widehat{U}^k(\mathbf{x}_p)~,
~~~~~~ p=1,2,\ldots,n.
\end{eqnarray}
where $\widehat{U}^k$ are the fictitious nodal nodes.

   Moreover, in order to uniquely determine $\widetilde{U}^k$, we also impose:

\begin{eqnarray}\label{constrain.poly}
\sum_{i=1}^{n}P_j(\mathbf{x}_i)a_i^k=0~,~~~~j=1,2,\ldots,m.
\end{eqnarray}

That is we have the following system of linear equations:

\begin{eqnarray}\nonumber
\mathbf{G}
\left[
 \begin{array}{c}
\mathbf{a}^k\\
\mathbf{b}^k\\
\end{array}
\right]=\left[\begin{array}{c}
\mathbf{\widehat{U}}^k\\
\mathbf{0}\\
\end{array}
\right]~,
\end{eqnarray}

where

\begin{eqnarray}\label{vectoru2}
\mathbf{\widehat{U}}^k=\left[
\begin{array}{cccc}
\widehat{U}_1^k & \widehat{U}_2^k & \ldots & \widehat{U}_n^k
\end{array}\right]^T = \left[
\begin{array}{cccc}
\widehat{U}^k(\mathbf{x}_1) & \widehat{U}^k(\mathbf{x}_2) & \ldots & \widehat{U}^k(\mathbf{x}_n)
\end{array}
\right]^T~,
\end{eqnarray}

\begin{eqnarray}\nonumber
\mathbf{G}=
\left[
 \begin{array}{cc}
\mathbf{R} & \mathbf{P}\\
\mathbf{P}^T & \mathbf{0}\\
\end{array}
\right],
\end{eqnarray}

 \begin{eqnarray}\nonumber
\mathbf{R} =
\left[
 \begin{array}{cccc}
R_1(\mathbf{x}_1) & R_2(\mathbf{x}_1) & \dots & R_n(\mathbf{x}_1)\\
R_1(\mathbf{x}_2) & R_2(\mathbf{x}_2) & \dots & R_n(\mathbf{x}_2)\\
\vdots     & \vdots     & \ddots   & \vdots   \\
R_1(\mathbf{x}_n) & R_2(\mathbf{x}_n) & \dots & R_n(\mathbf{x}_n)\\
\end{array}
\right]~,
\end{eqnarray}

 \begin{eqnarray}\nonumber
\mathbf{P} =
\left[
 \begin{array}{ccccccc}
P_1(\mathbf{x}_1) & P_2(\mathbf{x}_1) &  \dots & P_m(\mathbf{x}_1)\\
P_1(\mathbf{x}_2) & P_2(\mathbf{x}_2) &  \dots & P_m(\mathbf{x}_2)\\
\vdots     & \vdots  &  \ddots   & \vdots   \\
P_1(\mathbf{x}_n) & P_2(\mathbf{x}_n) & \dots & P_m(\mathbf{x}_n)\\
\end{array}
\right]~,
\end{eqnarray}

\begin{eqnarray}\label{vectora2}
\begin{array}{cccc}
\mathbf{a}^k=[a_1^k &a_2^k &\ldots& a_n^k]^T~,
\end{array}
\end{eqnarray}

\begin{eqnarray}\label{vectorb}
\begin{array}{cccc}
\mathbf{b}^k=[b_1^k &b_2^k &\ldots& b_m^k]^T~,
\end{array}
\end{eqnarray}

Unique solution is obtained if the inverse of matrix $\mathbf{R}$ exists, so that
\begin{eqnarray}\nonumber
\left[
 \begin{array}{c}
\mathbf{a}^k\\
\mathbf{b}^k\\
\end{array}
\right]
=\mathbf{G}^{-1}
\left[
 \begin{array}{c}
\mathbf{\widehat{U}}^k\\
\mathbf{0}\\
\end{array}
\right]~.
\end{eqnarray}

Accordingly, (\ref{RPI}) can be rewritten as

\begin{eqnarray}\nonumber
\widetilde{U}^k(\showx)=
\left[
 \begin{array}{cc}
\mathbf{R}^{T}(\mathbf{x}) & \mathbf{P}^{T}(\mathbf{x})\\
\end{array}
\right]
\left[
 \begin{array}{c}
\mathbf{a}^k\\
\mathbf{b}^k\\
\end{array}
\right]
~,
\end{eqnarray}

 or, equivalently,

\begin{eqnarray}\label{appr.rpi0}
\widetilde{U}^k(\mathbf{x})=
\left[
 \begin{array}{cc}
\mathbf{R}^{T}(\mathbf{x}) & \mathbf{P}^{T}(\mathbf{x})\\
\end{array}
\right]
\mathbf{G}^{-1}
\left[
 \begin{array}{c}
\mathbf{\widehat{U}}^k\\
\mathbf{0}\\
\end{array}
\right]
~.
\end{eqnarray}

Let us define the vector of shape functions:

\begin{eqnarray}\nonumber
\mathbf{\Phi}(\mathbf{x})=
[\begin{array}{cccccc}
\varphi_1(\mathbf{x}) & \varphi_2(\mathbf{x}) &\dots& \varphi_n(\mathbf{x})]
\end{array}
~,
\end{eqnarray}

where

\begin{eqnarray}\label{shaperb}
\varphi_p(\mathbf{x})=\sum_{i=1}^{n}R_{i}(\mathbf{x})\mathbf{G}^{-1}_{i,p}+\sum_{j=1}^{m}P_j(\mathbf{x})\mathbf{G}^{-1}_{n+j,p}
~, ~~~~p=1,2,\ldots,n~,
\end{eqnarray}
and $\mathbf{G}^{-1}_{i,p}$ is the $(i,p)$ element of the matrix $\mathbf{G}^{-1}$. \par
Using (\ref{shaperb}) relations (\ref{appr.rpi0}) are rewritten in the more compact form:

\begin{eqnarray}\label{appr.rpi2}
\widetilde{U}^k(\mathbf{x})=
\mathbf{\Phi}(\mathbf{x})\mathbf{\widehat{U}}^k
~,
\end{eqnarray}

or, equivalently,

\begin{eqnarray}\label{appr.rpi23}
\widetilde{U}^k(\mathbf{x})= \sum_{i=1}^{n} \widehat{U}_i^k \varphi_i(\mathbf{x})
~.
\end{eqnarray}


It can be easily shown that the shape functions (\ref{shaperb})
 satisfy the so-called Kronecker property, that is

\begin{eqnarray}\label{delta}
\varphi_i(\mathbf{x}_j) = \delta_{ij}~,
\end{eqnarray}

where $\delta_{ij}$ is the well-known Kronecker symbol,
 so that essential boundary and final conditions such as those
 considered in Section \ref{model} (e.g., relation (\ref{eq.main}))
 can be easily imposed.
Note also that the derivatives of $\widetilde{U}^k$ (of any order) with respect to $x$ or $z$  are easily obtained by direct differentiation in (\ref{appr.rpi23}).



\subsection{Discretized equations}\label{Dis.eq}
Before we show how to discretize model in the form (\ref{finalweak}), we focus on how to select nodal points.
Let $X=\{\mathbf{x}_0,\mathbf{x}_1,...,\mathbf{x}_N\}\subset\Omega$ are scattered meshless points, where some points are located on the boundary to enforce the boundary conditions. In fact, $\mathbf{x}_0$, $\mathbf{x}_N~\in \partial \Omega$. The options' payoffs considered in this paper are non-smooth functions, in particular
 their derivatives are discontinuous at the strike price. Therefore, to reduce the losses of accuracy the points of the trial functions are concentrated in a spatial region close to $s=E$. So, we satisfy this problem using relation (\ref{strans}) and the
following uniform nodal points along the $x$ and the $z$ directions, respectively:
\begin{eqnarray}\label{nodes}
&&x_i=i\Delta x,~~~~~~~i=0,1,...,N_x,\\\nonumber
&&z_j=j \Delta z,~~~~~~~j=0,1,...,N_z,
\end{eqnarray}
where $\Delta x=1/N_x$, $\Delta z=1/N_z$ and $N=(N_x+1)(N_z+1)$.
It is important to observe that $U^{k+1}(\mathbf{x})$ must be considered as known quantities, since it is approximated at the previous iteration.
We want to approximate $U^k(\mathbf{x})$ using LRPI approximation.
In the MLPG scheme, it is easy to enforce the boundary conditions (\ref{baghorban}) for that the shape function constructed by the LRPI approximation.
The LRPI approximation has shape functions with delta function properties, thus it allows one to easily impose
essential boundary and initial (or final) conditions.
\par
Substituting the displacement expression in Eq. (\ref{appr.rpi2})
into the local weak form (\ref{finalweak}) for each interior node in $\Omega_s^i$ the matrix forms of the their discrete equations are obtained as follows
 \begin{eqnarray}\label{sys.final}
\mathbf{F}\mathbf{\widehat{U}}^{k}=\mathbf{G}\mathbf{\widehat{U}}^{k+1},
\end{eqnarray}
where
\begin{eqnarray}\label{sysdast}
&&\mathbf{\widehat{U}}^{k} = [
\begin{array}{cccccccc}
\widehat{U}_{N_z+1}^{k} & \widehat{U}_{N_z+2}^{k} &\widehat{U}_2^{k} & \ldots& \widehat{U}_{N-N_z-1}^{k}
\end{array}
]^T_{(N-2N_z-1)\times1}.
\end{eqnarray}
Again we should note that in relation (\ref{sysdast}), $\widehat{U}_{0}^{k}$, $\widehat{U}_{1}^{k},~...,~\widehat{U}_{N_z}^{k}$ and $\widehat{U}_{N-N_z}^{k},~\widehat{U}_{N-N_z+1}^{k},~...,~\widehat{U}_{N}^{k}$ are calculated using the delta function properties easily.
Also in the linear system (\ref{sys.final}), $\mathbf{G}=[\mathbf{G}_{N_z+1}~~\mathbf{G}_{N_z+2}~~...~~\mathbf{G}_{N-N_z-1}]^T$ is the $(N-2N_z-1)\times(N-2N_z-1)$ banded matrix with bandwith $\texttt{bw}$ such that
\begin{eqnarray}\nonumber
\mathbf{G}_i=\frac{1}{\Delta t}\mathbf{\widetilde{E}}_{i}+\sum_{l=0}^{N}{}_{l}\mathbf{\widetilde{L}}_{i},~~~i=N_z+1,...,N-N_z-1,
\end{eqnarray}
where
\begin{eqnarray}\label{picew}
\{\mathbf{\widetilde{E}}_{i}\}_j=
\begin{cases}
\mathbf{\widetilde{E}}_{ij},& \mathbf{x}_j \in X \cap \Omega_s^i,\\
0,&o.w.
\end{cases}
\end{eqnarray}
This piecewise function which is defined to $\{\mathbf{\widetilde{E}}_{i}\}$, is extensible to ${}_{l}\mathbf{\widetilde{L}}_{i}$.
Also $\mathbf{F}=[\mathbf{F}_{N_z+1}~~\mathbf{F}_{N_z+2}~~...~~\mathbf{F}_{N-N_z-1}]^T$ is the $(N-2N_z-1)\times(N-2N_z-1)$ banded matrix with bandwith $\texttt{bw}$. We have
\begin{eqnarray}\label{Pi}
\mathbf{F}_i=\mathbf{\widetilde{A}}_{i}+\mathbf{\widetilde{B}}_{i}+\mathbf{\widetilde{C}}_{i}+\mathbf{\widetilde{D}}_{i}+(r+\lambda+\frac{1}{\Delta t})\mathbf{\widetilde{E}}_{i},~~~i=N_z+1,...,N-N_z-1.
\end{eqnarray}
Again we do emphasize that the piecewise function which is defined to $\{\mathbf{\widetilde{E}}_{i}\}$ in relation (\ref{picew}), is extensible to $\mathbf{\widetilde{A}}_{i},~\mathbf{\widetilde{B}}_{i},~\mathbf{\widetilde{C}}_{i}$ and $\mathbf{\widetilde{D}}_{i}$. Also we can easily see that
\begin{eqnarray}\nonumber
&&\mathbf{\widetilde{A}}_{ij}=\int_{\Omega_s^i}\mathbf{M}(\mathbf{x})\varphi_{j}(\mathbf{x})d\Omega,~~~~~~~~\mathbf{\widetilde{B}}_{ij}=\int_{\partial\Omega_s^i}\mathbf{N}(\mathbf{x})\varphi_{j}(\mathbf{x})d\Gamma,\\\nonumber
&&\mathbf{\widetilde{C}}_{ij}=\int_{\partial\Omega_s^i}\mathbf{I}(\mathbf{x})\frac{\partial}{\partial x}\varphi_{j}(\mathbf{x})d\Gamma,~~~~~~~\mathbf{\widetilde{D}}_{ij}=\int_{\partial\Omega_s^i}\mathbf{\Theta}(\mathbf{x})\frac{\partial}{\partial z}\varphi_{j}(\mathbf{x})d\Gamma,\\\nonumber
&&{}_{l}\mathbf{\widetilde{L}}_{ij}=\lambda\int_{\Omega_s^i}\int_{\Omega_s^l}\frac{r'(\hat{r})}{s(x)}f(\frac{\hat{r}}{s(x)})\varphi_{j}(\hat{r},z)~d\hat{r}~d\Omega,\\\nonumber
&&\mathbf{\widetilde{E}}_{ij}=\int_{\Omega_s^i}\varphi_{j}(\mathbf{x})d\Omega,\\\nonumber
\end{eqnarray}
where
\begin{eqnarray}\nonumber
&&\mathbf{M}(\mathbf{x})=(r-q-\lambda\kappa)\Bigg[\frac{s(x)}{s'(x)}\Bigg]'+\xi \Bigg[\frac{\eta-y(z)}{y'(z)}\Bigg]'+\frac{1}{2}y(z)\Bigg[\frac{s^2(x)s''(x)}{(s'(x))^3}\Bigg]-y(z)\Bigg[\frac{s(x)}{s'(x)}\Bigg]'\\\nonumber
&&~~~~~-\frac{\theta^2}{2}\frac{1}{y'(z)}+\frac{1}{2}\theta^2\frac{y(z)y''(z)}{(y'(z))^3}-\Bigg[\frac{s(x)}{s'(x)}\Bigg]\Bigg[\frac{y(z)}{y'(z)}\Bigg]',\\\nonumber
&&\mathbf{N}(\mathbf{x})=-(r-q-\lambda\kappa)\frac{s(x)}{s'(x)}\nu^1-\xi\frac{\eta-y(z)}{y'(z)}\nu^2+\frac{s(x)}{s'(x)}\Bigg[\frac{y(z)}{y'(z)}\Bigg]'\nu^1
+\frac{1}{2}y(z)\frac{s^2(x)}{s'(x)}\Bigg[\frac{1}{s'(x)}\Bigg]'\nu^1+y(z)\frac{s(x)}{s'(x)}\nu^1\\\nonumber
&&~~~~~+\frac{1}{2}\theta^2\frac{1}{y'(z)}\nu^2
+\frac{1}{2}\frac{y(z)}{y'(z)}\Bigg[\frac{1}{y'(z)}\Bigg]'\nu^2,\\\nonumber
&&\mathbf{I}(\mathbf{x})=-\frac{1}{2}y(z)\Bigg[\frac{s(x)}{s'(x)}\Bigg]^2\nu^1-\frac{s(x)y(z)}{s'(x)y'(z)}\nu^2,\\\nonumber
&&\mathbf{\Theta}(\mathbf{x})=-\frac{1}{2}\theta^2\frac{y(z)}{(y'(z))^2}\nu^2,\\\nonumber
\end{eqnarray}
Finally, combining Eqs.
(\ref{eq.final.amer.tran}) and (\ref{sys.final}) lead to the
following system:
\begin{eqnarray}\label{baz1.algorithm2}
\begin{cases}
\mathbf{F}\mathbf{\widehat{\Xi}}^{k}=\mathbf{G} \mathbf{\widehat{U}}^{k+1}~,\\
\mathbf{\widehat{U}}^{k}=\max\{\mathbf{\widehat{\Xi}}^{k},\mathbf{\widehat{\Pi}}\}~,
\end{cases}
\end{eqnarray}
to be recursively solved for $k=M-1,M-2,\ldots,0$, starting from
\begin{eqnarray}\label{baz1.final}
\mathbf{\widehat{U}}^{M} = \mathbf{\widehat{\Pi}},
\end{eqnarray}
  where $\mathbf{\widehat{\Pi}}$ are obtained from delta function properties of LRPI approximation and option's payoff (\ref{payoff.tran}).
\par
\textbf{Remark~2:} The numerical method proposed in this work
require solving at every time step a system of linear equations
(systems (\ref{sys.final})). Now, the matrix $\mathbf{F}$
associated to this system is band with bandwidth $\texttt{bw}$ and well-conditioned, therefore, the
aforementioned linear system is solved using the band LU
factorization method with partial pivoting, which is particularly
suitable for banded matrices. It should also be noted that the
complexity of banded LU factorization method with partial pivoting
is $\mathcal{O}(2N(2\texttt{bw}+3)(2\texttt{bw}+5))$. We simply
observe that complexity of this algorithm is very lower than
complexity of LU factorization method with partial pivoting for
strong form of MLPG or global RBF method which is
$\mathcal{O}(N^3/3)$. Moreover, as the matrix $\mathbf{F}$ to be
inverted are the same for every time step, the band LU
factorization can be performed only once at the beginning of the
numerical simulation, and thus at each time step the corresponding
linear system is efficiently solved by forward and backward
recursion (see \cite{G.H.Golub.book}).
\par
\textbf{Remark~3:} A crucial point in the MLPG is an accurate
evaluation of the local integrals. Since the nodal trial functions
based on LRPI are highly complicated, an accurate numerical
integration of the weak form is highly difficult. In this work,
the numerical integration procedure used is 4 points
Gauss-Legendre quadrature rule by a suitable change of variables.
\par
\subsection{Stability analysis}
In this section, we present an analysis of the stability of the presented scheme.
At first, we provide a new and simple notation
for $\mathbf{\widehat{U}}^{k}$, $\mathbf{F}$ and $\mathbf{G}$
\begin{eqnarray}\nonumber
&&\mathbf{\widehat{U}}^{k} =
[
\begin{array}{cccccccc}
\widehat{\mathcal{U}}_{0}^{k} & \widehat{\mathcal{U}}_{1}^{k} & \ldots& \widehat{\mathcal{U}}_{q}^{k}
\end{array}
]^T
=
[
\begin{array}{cccccccc}
\widehat{U}_{N_z+1}^{k} & \widehat{U}_{N_z+2}^{k} & \ldots& \widehat{U}_{N-N_z-1}^{k}
\end{array}
]^T_{(N-2N_z-1)\times1},\\\nonumber
&&\mathbf{F}=[\mathbf{\mathcal{F}}_{0}~~\mathbf{\mathcal{F}}_{1}~~...~~\mathbf{\mathcal{F}}_{q}]^T=[\mathbf{F}_{N_z+1}~~\mathbf{F}_{N_z+2}~~...~~\mathbf{F}_{N-N_z-1}]^T,\\\nonumber
&&\mathbf{G}=[\mathbf{\mathcal{G}}_{0}~~\mathbf{\mathcal{G}}_{1}~~...~~\mathbf{\mathcal{G}}_{q}]^T=[\mathbf{G}_{N_z+1}~~\mathbf{G}_{N_z+2}~~...~~\mathbf{G}_{N-N_z-1}]^T,
\end{eqnarray}
where $q=N-2N_z-2$.
In this scheme, the solution at any time level can be obtained using Eqs. (\ref{appr.rpi2}) and (\ref{baz1.algorithm2})
\begin{eqnarray}\label{stable1}
\mathbf{\widetilde{U}}^{k}=\mathbf{\phi}\max\{\mathbf{F}^{-1}\mathbf{G}\mathbf{\phi}^{-1}\mathbf{\widetilde{U}}^{k+1},\mathbf{\widetilde{\Pi}}\}~,
\end{eqnarray}
where $\mathbf{\phi}$ is the $(N-2N_z-1)\times(N-2N_z-1)$ identity matrix that
\begin{eqnarray}\nonumber
\mathbf{\widetilde{U}}^{k}=\mathbf{\phi}\mathbf{\widehat{U}}^{k},
\end{eqnarray}
and also we have
\begin{eqnarray}\nonumber
\mathbf{\widetilde{\Pi}}^{k}=\mathbf{\phi}\mathbf{\widehat{\Pi}}^{k}.
\end{eqnarray}
By choosing $k=l$ and using (\ref{stable1}), we get $\mathbf{\widetilde{U}}^{l}$. Assume that
\begin{eqnarray}\nonumber
&&\mathbf{\widehat{U}}^{l} = [
\begin{array}{cccccccc}
\widehat{\mathcal{U}}_0^{l} & \widehat{\mathcal{U}}_1^{l} &\widehat{\mathcal{U}}_2^{l} & \ldots& \widehat{\mathcal{U}}_q^{l}
\end{array}
]^T,\nonumber
\end{eqnarray}
Also, let $\mathbf{U}^{l}_{e}$ be the exact solution at the $l$th time level with the following components
\begin{eqnarray}\nonumber
&&\mathbf{U}^{l}_{e} = [
\begin{array}{cccccccc}
\mathcal{U}_{e0}^{l} & \mathcal{U}_{e1}^{l} &\mathcal{U}_{e2}^{l} & \ldots& \mathcal{U}_{eq}^{l}
\end{array}
]^T,\nonumber
\end{eqnarray}
It is well-known that for any $i=0,1,...,N$, $\widetilde{\mathcal{U}}^{l}_{i}$ is either less than $\mathcal{U}^{l}_{ei}$ or greater than it i.e.
\begin{eqnarray}\nonumber
\widetilde{\mathcal{U}}^{l}_{i}<\mathcal{U}^{l}_{ei},~~~~~~~~~\mathrm{or} ~~~~~~~~~~\widetilde{\mathcal{U}}^{l}_{i}\geq \mathcal{U}^{l}_{ei},~~~~~~~\forall ~i=0,1,..,q,
\end{eqnarray}
\textbf{Case 1}: Firstly, we consider the vector components of $\mathbf{\widetilde{U}}^{l}$ and $\mathbf{U}^{l}_{e}$ which have the following property
\begin{eqnarray}\nonumber
\widetilde{\mathcal{U}}^{l}_{i}\geq \mathcal{U}^{l}_{ei},
\end{eqnarray}
let us define the vectors ${}_{1}\mathbf{\widetilde{U}}^{l}$ and ${}_{1}\mathbf{U}^{l}_{e}$ as follow
\begin{eqnarray}\nonumber
{}_{1}\widetilde{\mathcal{U}}^{l}_{i}=
\begin{cases}
\widetilde{\mathcal{U}}^{l}_{i},&\widetilde{\mathcal{U}}^{l}_{i}\geq \mathcal{U}^{l}_{ei},\\
0,& o.w.
\end{cases}~~~~~~~~~~~~~
{}_{1}\mathcal{U}^{l}_{ei}=
\begin{cases}
\mathcal{U}^{l}_{ei},&\widetilde{\mathcal{U}}^{l}_{i}\geq \mathcal{U}^{l}_{ei},\\
0,& o. w.
\end{cases}
\end{eqnarray}
relation (\ref{stable1}) can be rewrite using the vector ${}_{1}\mathbf{\widetilde{U}}^{l}$ as follows
\begin{eqnarray}\label{bazgashticase1}
{}_{1}\mathbf{\widetilde{U}}^{l}=\mathbf{\phi}\max\{\mathbf{F}^{-1}\mathbf{G}\mathbf{\phi}^{-1}{}_{1}\mathbf{\widetilde{U}}^{l+1},\mathbf{M}\mathbf{\widetilde{\Pi}}\}~,
\end{eqnarray}
where $\mathbf{M}$ is a $(N-2N_z-1)\times(N-2N_z-1)$ matrix
\begin{eqnarray}
\mathbf{M}_{ij}=
\begin{cases}
1,& i=j,~~\mathrm{and}~~ \widetilde{\mathcal{U}}^{l}_{i}\geq \mathcal{U}^{l}_{ei},\\
0,& o. w.
\end{cases}
\end{eqnarray}
The error $\mathbf{E}_1^l$ at the $l$th time level is given by
\begin{eqnarray}
\mathbf{E}_1^l={}_{1}\mathbf{\widetilde{U}}^{l}- {}_{1}\mathbf{U}^{l}_{e},
\end{eqnarray}
It is important to observe that all components of $\mathbf{E}_1^l$ are positive values, also we conclude
\begin{eqnarray}\label{e1}
{}_{1}\mathbf{\widetilde{U}}^{l}=\mathbf{E}_1^l+{}_{1}\mathbf{U}^{l}_{e}.
\end{eqnarray}
Using the relations (\ref{bazgashticase1}) and (\ref{e1}), we get
\begin{eqnarray}\label{bazgashticase11}
\mathbf{E}_1^l+{}_{1}\mathbf{U}^{l}_{e}=\mathbf{\phi}\max\{\mathbf{F}^{-1}\mathbf{G}\mathbf{\phi}^{-1}{}_{1}\mathbf{U}^{l+1}_{e}+\mathbf{F}^{-1}\mathbf{Q}\mathbf{\phi}^{-1}\mathbf{E}_1^{l+1},\mathbf{M}\mathbf{\widetilde{\Pi}}\}~,
\end{eqnarray}
we can easily see that the relation (\ref{bazgashticase11}) is converted to the following equation using the maximum function properties:
\begin{eqnarray}\label{bazgashticase12}
\mathbf{E}_1^l+{}_{1}\mathbf{U}^{l}_{e}\leq\mathbf{\phi}\max\{\mathbf{F}^{-1}\mathbf{G}\mathbf{\phi}^{-1}{}_{1}\mathbf{U}^{l+1}_{e},\mathbf{M}\mathbf{\widetilde{\Pi}}\}+\mathbf{\phi}\max\{\mathbf{F}^{-1}\mathbf{G}\mathbf{\phi}^{-1}\mathbf{E}_1^{l+1},\mathbf{O}\}~,
\end{eqnarray}
where $\mathbf{O}$ is the zero vector. Also we know that
\begin{eqnarray}\label{bazgashticase13}
{}_{1}\mathbf{U}^{l}_{e}=\mathbf{\phi}\max\{\mathbf{F}^{-1}\mathbf{G}\mathbf{\phi}^{-1}{}_{1}\mathbf{U}^{l+1}_{e},\mathbf{M}\mathbf{\widetilde{\Pi}}\}~,
\end{eqnarray}
Therefore, using (\ref{bazgashticase12}) and (\ref{bazgashticase13}) we can write
\begin{eqnarray}
\mathbf{E}_1^l\leq\mathbf{\phi}\max\{\mathbf{F}^{-1}\mathbf{G}\mathbf{\phi}^{-1}\mathbf{E}_1^{l+1},\mathbf{O}\}~,
\end{eqnarray}
Finally, we obtain
\begin{eqnarray}
||\mathbf{E}_1^l||\leq||\mathbf{\phi}\max\{\mathbf{F}^{-1}\mathbf{G}\mathbf{\phi}^{-1}\mathbf{E}_1^{l+1},\mathbf{O}\}||\leq||\mathbf{\phi}\mathbf{F}^{-1}\mathbf{G}\mathbf{\phi}^{-1}\mathbf{E}_1^{l+1}||\leq||\mathbf{\phi}\mathbf{F}^{-1}\mathbf{G}\mathbf{\phi}^{-1}||||\mathbf{E}_1^{l+1}||~,
\end{eqnarray}
or, equivalently
\begin{eqnarray}
||\mathbf{E}_1^l||\leq||\mathbf{\phi}\mathbf{F}^{-1}\mathbf{G}\mathbf{\phi}^{-1}||||\mathbf{E}_1^{l+1}||~,
\end{eqnarray}
\textbf{Case 2}. Now, we consider the vector components of $\mathbf{\widetilde{U}}^{l}$ and $\mathbf{U}^{l}_{e}$ which have the following property
\begin{eqnarray}\nonumber
\widetilde{\mathcal{U}}^{l}_{i}< \mathcal{U}^{l}_{ei},
\end{eqnarray}
suppose that ${}_{2}\mathbf{\widetilde{U}}^{l}$ and ${}_{2}\mathbf{U}^{l}_{e}$ are two vectors defined by
\begin{eqnarray}\nonumber
{}_{2}\widetilde{\mathcal{U}}^{l}_{i}=
\begin{cases}
\widetilde{\mathcal{U}}^{l}_{i},&\widetilde{\mathcal{U}}^{l}_{i}< \mathcal{U}^{l}_{ei},\\
0,& o.w.
\end{cases}~~~~~~~~~~~~~
{}_{2}\mathcal{U}^{l}_{ei}=
\begin{cases}
\mathcal{U}^{l}_{ei},&\widetilde{\mathcal{U}}^{l}_{i}< \mathcal{U}^{l}_{ei},\\
0,& o. w.
\end{cases}
\end{eqnarray}
Anyway, another alternative for Eq. (\ref{stable1}) related to ${}_{2}\mathbf{\widetilde{U}}^{l}$ can be mentioned as
\begin{eqnarray}\label{baz2}
{}_{2}\mathbf{\widetilde{U}}^{l}=\mathbf{\phi}\max\{\mathbf{F}^{-1}\mathbf{G}\mathbf{\phi}^{-1}{}_{2}\mathbf{\widetilde{U}}^{l+1},\mathbf{N}\mathbf{\widetilde{\Pi}}\}~,
\end{eqnarray}
where $\mathbf{N}$ is a $(N-2N_z-1)\times(N-2N_z-1)$ matrix defined by
\begin{eqnarray}
\mathbf{N}_{ij}=
\begin{cases}
1,& i=j,~~\mathrm{and}~~ \widetilde{\mathcal{U}}^{l}_{i}< \mathcal{U}^{l}_{ei},\\
0,& o. w.
\end{cases}
\end{eqnarray}
In this case we propose the Error $\mathbf{E}_2^l$ at the $l$th time level
\begin{eqnarray}\label{e2}
\mathbf{E}_2^l={}_{2}\mathbf{\widetilde{U}}^{l}- {}_{2}\mathbf{U}^{l}_{e},
\end{eqnarray}
It is clear that $\mathbf{E}_2^l$ is hold as
\begin{eqnarray}
\mathbf{E}_2^l\geq0,
\end{eqnarray}
By using relation (\ref{e2}), we obtain
\begin{eqnarray}
{}_{2}\mathbf{\widetilde{U}}^{l}={}_{2}\mathbf{U}^{l}_{e}-\mathbf{E}_2^l.
\end{eqnarray}
Therefore, the relation (\ref{baz2}) converted to the following equation
\begin{eqnarray}
{}_{2}\mathbf{U}^{l}_{e}-\mathbf{E}_2^l=\mathbf{\phi}\max\{\mathbf{F}^{-1}\mathbf{G}\mathbf{\phi}^{-1}{}_{2}\mathbf{U}^{l+1}_{e}-\mathbf{F}^{-1}\mathbf{G}\mathbf{\phi}^{-1}\mathbf{E}_2^{l+1},\mathbf{N}\mathbf{\widetilde{\Pi}}\}~,
\end{eqnarray}
Moreover, using the maximum function property, we have
\begin{eqnarray}
{}_{2}\mathbf{U}^{l}_{e}-\mathbf{E}_2^l\geq\mathbf{\phi}\max\{\mathbf{F}^{-1}\mathbf{G}\mathbf{\phi}^{-1}{}_{2}\mathbf{U}^{l+1}_{e},\mathbf{N}\mathbf{\widetilde{\Pi}}\}-\mathbf{\phi}\max\{\mathbf{F}^{-1}\mathbf{G}\mathbf{\phi}^{-1}\mathbf{E}_2^{l+1},\mathbf{O}\}~,
\end{eqnarray}
or
\begin{eqnarray}
0\leq\mathbf{E}_2^l\leq\mathbf{\phi}\max\{\mathbf{F}^{-1}\mathbf{G}\mathbf{\phi}^{-1}\mathbf{E}_2^{l+1},\mathbf{O}\}~,
\end{eqnarray}
Then, it follows from the norm and maximum property that
\begin{eqnarray}
||\mathbf{E}_2^l||\leq||\mathbf{\phi}\max\{\mathbf{F}^{-1}\mathbf{G}\mathbf{\phi}^{-1}\mathbf{E}_2^{l+1},\mathbf{O}\}||\leq||\mathbf{\phi}\mathbf{F}^{-1}\mathbf{G}\mathbf{\phi}^{-1}\mathbf{E}_2^{l+1}||\leq||\mathbf{\phi}\mathbf{F}^{-1}\mathbf{G}\mathbf{\phi}^{-1}||||\mathbf{E}_2^{l+1}||~,
\end{eqnarray}
or, equivalently
\begin{eqnarray}
||\mathbf{E}_2^l||\leq||\mathbf{\phi}\mathbf{F}^{-1}\mathbf{G}\mathbf{\phi}^{-1}||||\mathbf{E}_2^{l+1}||~.
\end{eqnarray}
The numerical scheme will be stable if $l\rightarrow \infty$, the error $||\mathbf{E}_1^l||\rightarrow0$ and $||\mathbf{E}_2^l||\rightarrow0$.
This can be guaranteed provided $\rho(\mathbf{\phi}\mathbf{F}^{-1}\mathbf{G}\mathbf{\phi}^{-1})\leq1$ or $\rho(\mathbf{F}^{-1}\mathbf{G})\leq1$ (because $\mathbf{F}^{-1}\mathbf{G}$ and $\mathbf{\phi}\mathbf{F}^{-1}\mathbf{G}\mathbf{\phi}^{-1}$ are similar matrices), where $\rho$ denoted the spectral radius of the matrix.\par
For the analysis, we need a simple version of the matrices $\mathbf{F}$ and $\mathbf{G}$. It is given by
\begin{eqnarray}\nonumber
&&\mathbf{F}=\mathbf{\widetilde{A}}
+\mathbf{\widetilde{B}}+\mathbf{\widetilde{C}}+\mathbf{\widetilde{D}}+(r+\lambda+\frac{1}{\Delta t})\mathbf{\widetilde{E}},\\\nonumber
&&\mathbf{G}=\frac{1}{\Delta t}\mathbf{\widetilde{E}}+\sum_{l=0}^{N}{}_{l}\mathbf{\widetilde{L}}
\end{eqnarray}
where $\mathbf{\widetilde{A}}$, $\mathbf{\widetilde{B}}$, $\mathbf{\widetilde{C}}$, $\mathbf{\widetilde{D}}$, $\mathbf{\widetilde{E}}$ and ${}_{l}\mathbf{\widetilde{L}}$ are the $(N-2N_z-1)\times(N-2N_z-1)$ sparse matrices whose the rows of them are obtain using relation (\ref{Pi}).
\par
\begin{figure}
\center
\includegraphics[width=17cm,height=11cm]{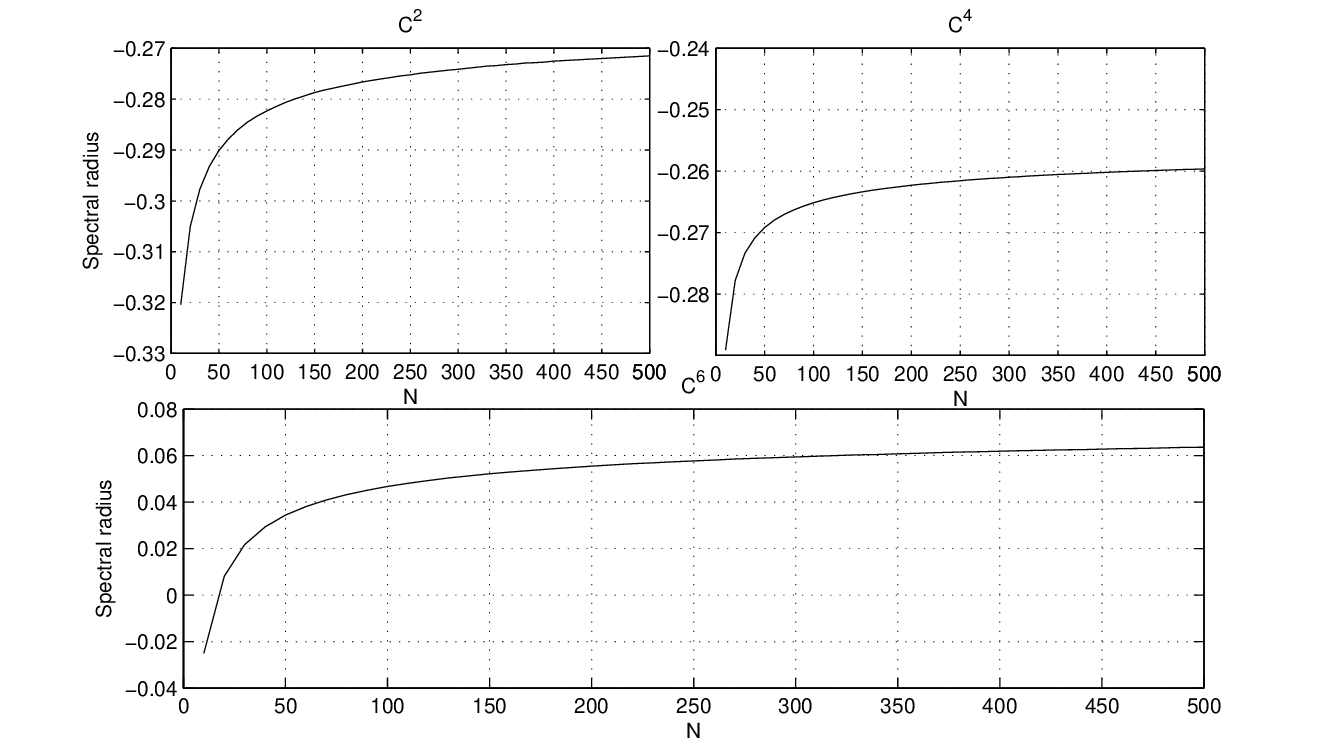}
\caption{Spectral radius of $\mathbf{\Sigma}^{-1}\mathbf{\Gamma}$ in LRPI methods based on Wendland's compactly supported radial basis functions (WCS-RBFs) with $C^6$
, $C^4$ and $C^2$ smoothness degrees.} \label{SrFig}
\end{figure}

Then we obtain
\begin{eqnarray}\label{pp}
&&\mathbf{F}=\mathbf{S} +\frac{1}{\Delta t}\mathbf{\widetilde{E}},\\\nonumber
&&\mathbf{G}=\mathbf{Q}+\frac{1}{\Delta t}\mathbf{\widetilde{E}},
\end{eqnarray}
where
\begin{eqnarray}\nonumber
&&\mathbf{S}=\mathbf{\widetilde{A}}
+\mathbf{\widetilde{B}}+\mathbf{\widetilde{C}}+\mathbf{\widetilde{D}}+(r+\lambda)\mathbf{\widetilde{E}},\\\nonumber
&&\mathbf{Q}=\sum_{l=0}^{N}{}_{l}\mathbf{\widetilde{L}}.
\end{eqnarray}
However, we can consider
\begin{eqnarray}\label{bazgasht}
&&\mathbf{\widetilde{E}}^{-1}\mathbf{F}=\mathbf{\widetilde{E}}^{-1}\mathbf{S} +\frac{1}{\Delta t}\mathbf{I},\\\nonumber
&&\mathbf{\widetilde{E}}^{-1}\mathbf{G}=\mathbf{\widetilde{E}}^{-1}\mathbf{Q}+\frac{1}{\Delta t}\mathbf{I},
\end{eqnarray}
on the other hand, we know that
\begin{eqnarray}\nonumber
\mathbf{F}^{~-1}\mathbf{G}=\mathbf{F}^{~-1}\mathbf{\widetilde{E}}~\mathbf{\widetilde{E}}^{~-1}\mathbf{G}=
(\mathbf{\widetilde{E}}^{~-1}\mathbf{F})^{~-1}(\mathbf{\widetilde{E}}^{~-1}\mathbf{G}),
\end{eqnarray}
let us define
\begin{eqnarray}\nonumber
\mathbf{\Sigma}=\mathbf{\widetilde{E}}^{-1}\mathbf{F},~~~~
\mathbf{\Gamma}=\mathbf{\widetilde{E}}^{-1}\mathbf{G},~~~~
\mathbf{\Upsilon}=\mathbf{\widetilde{E}}^{-1}\mathbf{S},~~~~
\mathbf{\Psi}=\mathbf{\widetilde{E}}^{-1}\mathbf{Q},~~~~
\end{eqnarray}
therefore, we can rewrite relation (\ref{bazgasht}) as follows
\begin{eqnarray}\label{bazgasht}
&&\mathbf{\Sigma}=\mathbf{\Upsilon}+\frac{1}{\Delta t}\mathbf{I},\\\nonumber
&&\mathbf{\Gamma}=\mathbf{\Psi}+\frac{1}{\Delta t}\mathbf{I},
\end{eqnarray}
Now by applying Cayley-Hamilton theorem and Gelfand's formula, we have
\begin{eqnarray}\label{stablefinal}
\mathbf{\rho}(\mathbf{\overline{F}}^{~-1}\mathbf{\overline{G}})=\mathbf{\rho}(\mathbf{\Sigma}^{-1}\mathbf{\Gamma})
\leq\Bigg|\frac{\Delta t\mathbf{\rho}(\mathbf{\Upsilon})+1}{\Delta t\mathbf{\rho}(\mathbf{\Psi})+1}\Bigg|<1,
\end{eqnarray}
where $\mathbf{\rho}$ is spectral radius of the matrices.
We can easily see that the inequality (\ref{stablefinal}) is
 always satisfied and the scheme will be unconditionally stable if $\rho(\mathbf{\Upsilon})\leq\mathbf{\rho}(\mathbf{\Psi})$. Figure \ref{SrFig} shows numerically how $\rho(\mathbf{\Upsilon})-\mathbf{\rho}(\mathbf{\Psi})$ varies as a function of $N$. Recollect that the stability condition is satisfied only when $\rho(\mathbf{\Upsilon})-\mathbf{\rho}(\mathbf{\Psi})\leq0$.
  It can be seen from Figure \ref{SrFig} that this condition is satisfied in the present numerical method.\par
%
\par
\section{Numerical results and discussions}
To get a better sense of the efficiency of the method presented
in the current paper, let us employ the scheme in solving some test problems.
Following the notation employed in Section 3, let $V$ and $V_{LRPI}$ respectively denote the option price (either European
or American) and its approximation obtained using the LRPI method developed in the previous section.
To measure the accuracy of the $V_{LRPI}$ method at the current time, the
discrete maximum norm and the root mean square relative
difference (RMSRD) have been used with
the following definitions:
\begin{eqnarray}\label{errormax}
&&\textrm{MaxError}_{LRPI} = \max_{i=0,1,\ldots,l}
\left|V_{LRPI}(S_i,y_0,0) - V(S_i,y_0,0) \right|  ,
\end{eqnarray}
\begin{eqnarray}\label{errorrms}
&&\textrm{RMSRD}_{LRPI} = \frac{1}{l+1}\sqrt{\sum_{i=0}^l
\left(\frac{ V_{LRPI}(S_i,y_0,0) - V(S_i,y_0,0)}{V(S_i,y_0,0)} \right)^2}  .
\end{eqnarray}
In $\textrm{MaxError}_{LRPI}$ and $\textrm{RMSRD}_{LRPI}$, $S_i,
~i=0,1,..,l$ are $l+1$ different points that will be chosen in a
convenient neighborhood of the strike $E$, i.e. $S_i \in
(\frac{4}{5}E,\frac{6}{5}E)$. For simplicity, in European and
American options we set $S_i=(0.1i+0.8)E$, where $i\in\Xi_1=\{0,1,2,3,4\}$ or $i\in\Xi_2=\{1,2,3\}$. Note that only in the case of
the European option under SV model the exact value of $V$ is available.
Therefore to the other methods we use instead the reference prices which are described in previous papers, where they have been obtained by
performing an accurate (and also very time-consuming) simulation on a very refined mesh.
\par
In the following analysis,
the optimal values of the radius of the local sub-domains is
selected using the figures
for $MaxError$ (or RMSRD) vs different values of $r_{Q}$ (see Figures \ref{Test1E} and \ref{Test1A} for SV model; Figures \ref{test2E}, \ref{test2Aa}, \ref{test2Ab}, \ref{test2Ac} and \ref{test2Ad} for SVJ model; and Figures \ref{test3E} and \ref{test3A} for SVCJ model).
The size of $r_{Q}$ is such that the union of these sub-domains must cover the whole global domain i.e. $\cup
\Omega_s^i\subset\Omega$. It is also worth noticing that the MLS
approximation is well-defined only when $\mathbf{G}$ is
non-singular or the rank of $\mathbf{P}$ equals $m$ and at least
$m$ radial functions are non-zero i.e. $n>m$ for each $\mathbf{x} \in
\Omega$. Therefore, to satisfy these conditions, the size of the
support domain $r_w$ should be large enough to have sufficient
number of nodes covered in $\Omega_s^i$ for every sample point
($n>m$). In all the simulations presented in this work we use
$r_w=l~h$, where $l=1.5,2,2.5,3$, and $h$ is the distance between the nodes.
Figures \ref{Test1E} and \ref{Test1A} for SV model; Figures \ref{test2E}, \ref{test2Aa}, \ref{test2Ab}, \ref{test2Ac} and \ref{test2Ad} for SVJ model; and Figures \ref{test3E} and \ref{test3A} for SVCJ model are considered to illustrate the effect of the radius of the local sub-domains $r_Q$ and the size of the
support domain $r_w$ on our solutions. In these figures, the effect of
$r_Q$ and $r_w$ on $MaxError$ (or RMSRD) are shown. The radius of the local sub-domains $r_Q$ and the size of the
support domain $r_w$ should be chosen to reduce the value
of $MaxError$ (or RMSRD). From these figures, it can be seen that in all the simulations presented in this work, the accuracy grows as the size of the
support domain $r_w$ increases gradually. On the
other hand, we know that an increase in the size of the
support domain $r_w$, increases the CPU time
computed in the approximation, and this is a fact that in this paper
the best value of the size of the
support domain $r_w$ is $1.5h$.\par
Also, for all models we use $\xi_s=1$ and $\xi_y=10$. These values are chosen by trial and
error such as to roughly minimize the errors on the numerical
solutions.
\par
To show the rate of convergence of the new scheme when
$h\rightarrow0$ and $\Delta t\rightarrow0$, the values of ratio
with the following formula have been reported in the tables
\begin{eqnarray}\nonumber
&&\mathrm{Ratio}_{LRPI}=\log_2\Bigg\{\frac{\textrm{MaxError}_{LRPI}~(\mathrm{or~RMSRD}_{LRPI}) ~~\textrm{in the previous row}}{\textrm{MaxError}_{LRPI}~(\mathrm{or~RMSRD}_{LRPI}) ~~\textrm{in the current row}}\Bigg\}.
\end{eqnarray}
Also, the computer time required to obtain the option price
using the numerical method described in previous section is
denoted by $CPU~time$.

Finally, the numerical implementation and all of the executions are performable by Matlab software, alongside hardware configuration: Intel(R) Core(TM)2 Duo CPU T9550  2.66 GHz
  4 GB RAM.\par

\subsection{Test case 1: SV models}
To demonstrate the excellent capability of the presented method, first example considers the European and American options under SV model. In
particular, we consider the same test case reported in \cite{Clarke,Oosterlee,Zvan.Forsyth,Salmi.N.M,Salmi.NMPDE,Persson,S.P.Zhu.Chen,ITO}, where the option
and model parameters are chosen as in Table \ref{Tab.par}.\par
\begin{table}[tbh]\tiny
\caption{Model parameters and data. (E.p: European put, A.p: American put, A.c: American call.)}
\begin{tabular*}{\columnwidth}{@{\extracolsep{\fill}}*{22}{c}}
\hline
&& $E$ & $T$ (year) & $r$  & $q$  & $\xi$  & $\eta$  & $\theta$& $\lambda$  & $\delta$ & $\gamma$ & $\rho$ & $\rho_j$ & $\nu$\\
\hline
Test case 1&&&&&&&&&&&&&&\\
& E.p \cite{Heston} & 10 & 0.25 & 0.1 & - & 5 & 0.16 & 0.9 & - & - & - & 0.1 & - & -\\
&A.p  \cite{Clarke,Oosterlee,Zvan.Forsyth,Salmi.N.M,Salmi.NMPDE,Persson,S.P.Zhu.Chen,ITO} & 10 & 0.25 & 0.1 & - & 5 & 0.16 & 0.9 & - & - & - & 0.1 & - & - \\
&&&&&&&&&&&&&&\\
Test case 2&&&&&&&&&&&&&&\\
&E.p \cite{Salmi}   & 100 & 0.5  & 0.03     & -    & 2       & 0.04   & 0.25 & 0.2 & 0.04  & -0.5   & -0.5 & - & - \\
&A.p \cite{Salmi}   & 100 & 0.5  & 0.03     & -    & 2       & 0.04   & 0.25 & 0.2 & 0.04  & -0.5   & -0.5 & - & - \\
&A.c \cite{Ballestra.4,Chiarella.Ziogas}   & 100 & 0.5  & 0.03     & 0.05 & 2       & 0.04   & 0.4  & 5   & 0.1  & -0.005 & 0.5  & - & - \\
&A.c \cite{Ballestra.4,Chiarella.Ziogas}   & 100 & 0.5  & 0.03     & 0.05 & 2       & 0.04   & 0.4  & 5   & 0.1  & -0.005 & -0.5 & - & - \\
&A.c \cite{Ballestra.4,Toivanen.Bal}   & 100 & 0.5  & 0.03     & 0.05 & 2       & 0.04   & 0.25 & 0.2 & 0.4  & -0.5   & -0.5 & - & - \\
&&&&&&&&&&&&&&\\
Test case 3&&&&&&&&&&&&&&\\
&E.p \cite{Salmi}   & 100 & 0.5  & 0.03     & -    & 2       & 0.04   & 0.25 & 0.2 & 0.04  & -0.5   & -0.5 & -0.5 & 0.2 \\
&A.p \cite{Salmi}   & 100 & 0.5  & 0.03     & -    & 2       & 0.04   & 0.25 & 0.2 & 0.04  & -0.5   & -0.5 & -0.5 & 0.2 \\
\end{tabular*}
\label{Tab.par}
\end{table}
\par
It should be noted that in all simulations proposed in this work, we have $[0,S_{max}]\times[0,y_{max}]=[0,4E]\times[0,1]$.
Again we do emphasize that in the American option under SV model, the exact value of $V$ is not available.
Therefore we use instead the reference prices which are described in \cite{Clarke,Oosterlee,Zvan.Forsyth,Salmi.N.M,Salmi.NMPDE,Persson,S.P.Zhu.Chen,ITO}.
The reference prices for these options are given in Table \ref{Tab.main.CV.A}.
\par
\begin{table}[tbh]\tiny
\caption{Reference prices used in American put option under SV model.}
\begin{tabular*}{\columnwidth}{@{\extracolsep{\fill}}*{10}{c}}
\hline
References& &($N_x,~N_z,~M$)& $y$ & Value at 8 & Value at 9 & Value at 10 & Value at 11  & Value at 12 & error\\
\hline
&&&&&&\\
Ikonen and Toivanen \cite{IkonenPSOR} & & (4096,2048,4098)  &&&&&&\\
&&& 0.0625 & 2.000000     & 1.107629   & 0.520038   & 0.213681   & 0.082046 \\
&&&                                0.25   & 2.078372     & 1.333640   & 0.795983   & 0.448277   & 0.242813  \\
&&&&&&&\\
Presented method   & & (2048,1024,1024)  &&&&&&&\\
&$C^2$&& 0.0625                           & 2.000003   & 1.107624   & 0.520036   & 0.213682   & 0.082049 &5.00E-06\\
&&&                                0.25   & 2.078378   & 1.333643   & 0.795984   & 0.448276   & 0.242809 &6.00E-06\\
&$C^4$&& 0.0625                           & 2.000000   & 1.107627   & 0.520038   & 0.213675   & 0.082042 &4.50E-06\\
&&&                                0.25   & 2.07837   & 1.333642   & 0.795981   & 0.448276   & 0.242811 &2.00E-06\\
&$C^6$&& 0.0625                           & 2.000000   & 1.107628   & 0.520038   & 0.21368   & 0.082044 &2.00E-06\\
&&&                                0.25   & 2.078372   & 1.333641   & 0.795981   & 0.448276   & 0.242811 &2.00E-06\\
&&&&&&&\\
Ito and Toivanen \cite{ITO}   & & (2049,1025,1025)  &&&&&&&\\
&&& 0.0625 & 2.000000   & 1.107621   & 0.520030   & 0.213677   & 0.082044 &8.00E-06\\
&&&                                0.25   & 2.078364   & 1.333632   & 0.795977   & 0.448273   & 0.242810 &8.00E-06\\
&&&&&&&\\
Clarke and Parrott \cite{Clarke} && (513,193,-) &&&&&&&\\
&&& 0.0625 & 2.0000     & 1.1080   & 0.5316   & 0.2261   & 0.0907 &1.24E-02\\
&&&                                0.25   & 2.0733     & 1.3290   & 0.7992   & 0.4536   & 0.2502 & 7.39E-03\\
&&&&&&&\\
Zvan, Forsyth \cite{Zvan.Forsyth}       && (177,103,-)   &&&&&&&\\
&&& 0.0625 & 2.0000     & 1.1076   & 0.5202   & 0.2138   & 0.0821 &1.62E-04\\
&&&                                0.25   & 2.0784     & 1.3337   & 0.7961   & 0.4483   & 0.2428 &1.17E-04\\
&&&&&&&\\
Oosterlee \cite{Oosterlee}    && (257,257,-)          &&&&&&&\\
&&& 0.0625 & 2.000     & 1.107   & 0.517   & 0.212   & 0.0815 &3.04E-03\\
&&&                                0.25   & 2.079     & 1.334   & 0.796   & 0.449   & 0.243   & 7.23E-04\\
&&&&&&&\\
Yousuf and Khaliq \cite{Yousuf.Khaliq.NMPDE.ETD}   & & (400,80,20)  &&&&&&&\\
&&& 0.0625 & 1.9958     & 1.1051   & 0.5167   & 0.2119   & 0.0815 &4.20E-03\\
&&&                                0.25   & 2.0760     & 1.3316   & 0.7945   & 0.4473   & 0.2423  & 2.37E-03\\
&&&&&&&\\
Ikonen and Toivanen \cite{Salmi.N.M} & & (320,128,64)  &&&&&&&\\
&&& 0.0625 & 2.00000     & 1.10749   & 0.51985   & 0.21354   & 0.08198 &1.88E-04\\
&&&                                0.25   & 2.07829     & 1.33351   & 0.79583   & 0.44815   & 0.24273  & 1.53E-04\\
&&&&&&&\\
\hline
\end{tabular*}
\label{Tab.main.CV.A}
\end{table}
\par
Testifying the accuracy, numerical rate of convergence of the solution and CPU time with respect to the number of scattered nodes are
of our special interest.
To achieve this goal, we apply the local weak form meshless method by employing the radial point interpolation based on Wendland's compactly
supported radial basis functions with $C^2$, $C^4$ and $C^6$ smoothness together with different choice of $N_x$, $N_z$ and $M$ to evaluate
European and American options of this financial model. The result are given in Tables \ref{Tab.test1.E} and \ref{Tab.test1.A}.
\par
\begin{table}[tbh]\tiny
\caption{Test case 1: European put option.}
\begin{tabular*}{\columnwidth}{@{\extracolsep{\fill}}*{11}{c}}
\hline
\multicolumn{1}{c}{}&\multicolumn{3}{c}{$C^2$}& \multicolumn{3}{c}{$C^4$}&\multicolumn{3}{c}{$C^6$}\\
\cline{2-4} \cline{5-7}\cline{8-10}
($N_x,~N_z,~M$) & error & Ratio & CPU time & error & Ratio & CPU time & error & Ratio & CPU time\\
\hline
&&&\\
(16,8,8)        & 4.30E-05 &-&0.00    & 2.78E-05 &-&0.00    & 2.11E-05 &-&0.00\\
(32,16,16)      & 1.91E-05 &1.17&0.00 & 7.70E-06 &1.85&0.00 & 7.93E-06 &1.41&0.00\\
(64,32,32)      & 5.55E-06 &1.78&0.00 & 2.03E-06 &1.92&0.00 & 2.02E-06 &1.97&0.00\\
(128,64,64)     & 1.72E-06 &1.69&0.19 & 5.37E-07 &1.92&0.19 & 5.40E-07 &1.90&0.19\\
(256,128,128)   & 4.81E-07 &1.84&0.42 & 1.38E-07 &1.96&0.42 & 1.36E-07 &1.99&0.42\\
(512,256,256)   & 1.28E-07 &1.91&1.80 & 3.48E-08 &1.99&1.80 & 3.39E-08 &2.00&1.80\\
(1024,512,512)  & 3.37E-08 &1.92&7.01 & 8.68E-09 &2.00&7.01 & 8.42E-09 &2.01&7.01\\
(2048,1024,1024)& 8.89E-09 &1.92&24.43& 2.16E-09 &2.01&24.43& 2.09E-09 &2.01&24.43\\
\hline
\end{tabular*}
\label{Tab.test1.E}
\end{table}
\par
\begin{table}[tbh]\tiny
\caption{Test case 1: American put option.}
\begin{tabular*}{\columnwidth}{@{\extracolsep{\fill}}*{11}{c}}
\hline
\multicolumn{1}{c}{}&\multicolumn{3}{c}{$C^2$}& \multicolumn{3}{c}{$C^4$}&\multicolumn{3}{c}{$C^6$}\\
\cline{2-4} \cline{5-7}\cline{8-10}
($N_x,~N_z,~M$) & error & Ratio & CPU time & error & Ratio & CPU time & error & Ratio & CPU time\\
\hline
&&&\\
(16,8,8)        & 7.30E-02 &-&0.00    & 5.16E-02 &-&0.00    & 1.87E-02 &-&0.00\\
(32,16,16)      & 7.89E-03 &3.21&0.00 & 1.69E-02 &1.61&0.00 & 5.30E-03 &1.82&0.00\\
(64,32,32)      & 2.48E-03 &1.67&0.00 & 4.67E-03 &1.86&0.00 & 1.43E-03 &1.89&0.00\\
(128,64,64)     & 1.03E-03 &1.27&0.33 & 1.21E-03 &1.95&0.33 & 3.83E-04 &1.90&0.33\\
(256,128,128)   & 2.68E-04 &1.94&0.81 & 3.07E-04 &1.98&0.81 & 1.02E-04 &1.91&0.81\\
(512,256,256)   & 7.09E-05 &1.92&2.70 & 7.73E-05 &1.99&2.70 & 3.13E-05 &1.70&2.70\\
(1024,512,512)  & 1.98E-05 &1.84&11.14& 1.92E-05 &2.01&11.14& 7.94E-06 &1.98&11.14\\
(2048,1024,1024)& 6.00E-06 &1.72&35.68& 4.50E-06 &2.09&35.68& 2.00E-06 &1.99&35.68\\
\hline
\end{tabular*}
\label{Tab.test1.A}
\end{table}
\par
Following these consequences, we derive the fast and accurate solutions that possess both properties convergence and stability from the numerical
point of view. The tables given in this model obviously confirm this claim.
The number
of time discretization steps is set equal to nodes distributed in
the volatility dimension of the asset price. As we have experimentally checked, this choice is such
that in all the simulations performed the error due to the time
discretization is negligible with respect to the error due to the
 LRPI discretization (note that in the present work we are
mainly concerned with the LRPI spatial approximation).
Paying attention, from these tables we observe that the accuracy grows as the number of nodes increases gradually, this fact
can be understood more clear from the data shown in Tables \ref{Tab.test1.E} and \ref{Tab.test1.A} since as the number of nodes increases, we find the approximations that possess more truly significant digits. Then the option price can be computed with a small financial error in a small computer time.
This indicates that the numerical solution converges to the true solution as the number of nodes increases gradually. Again we do confirm that the true solution is available only in European option which is proposed in \cite{Heston}. Therefore , in American option, we use instead the reference price which is described in \cite{IkonenPSOR}, where it has been obtained by
performing an accurate (and also very time-consuming) simulation on a very refined mesh.
\par
In fact, considering Table \ref{Tab.test1.E}, by employing $64\times32$ nodal points in domain and $32$ time discretization steps, European option under SV model is computed with an error of
 order $O(10^{-6})$ in $0$ second, instead, using $128\times64$ nodal points and $64$ time discretization steps,
 the option price is computed with an error of order $O(10^{-7})$ in $0.19$ second and using $512\times256$ nodal points and $256$ time discretization steps,
 the option price is computed with an error of order $O(10^{-8})$ in 1.80 second.
 This means especially that the computer times necessary to perform these simulations are extremely small.
 Note that, for example, in Table \ref{Tab.test1.E}, as well as in the following ones, the fact that error is approximately
 $10^{-6}$ means that $V_{LRPI}(s,y_0,0)$ is up to at least the 6th significant digit, equal to $V(s,y_0,0)$.
On the other hand, by looking at Table \ref{Tab.test1.A}, it can also be
seen that in LRPI scheme error of orders $O(10^{-3}),~O(10^{-4})$ and
$O(10^{-5})$ are computed using $64\times32$, $256\times128$ and $512\times256$ nodal points and $32,~128,~256$
time discretization steps, respectively in 0, 0.33 and 0.81
seconds.
As the additional point, we can observe that the option price computed by $C^2$, $C^4$ and $C^6$ Wendland's
compactly supported radial basis functions on
sub-domains require the same very small CPU times to reach an accuracy
of $6\times 10^{-6}$, $4.5\times 10^{-6}$ and $2\times10^{-6}$ on the same number of
sub-domains.
Therefore, we can simply conclude that in Wendland's
compactly supported radial basis functions,  the accuracy grows as the smoothness order of these functions increases.
The next issue is examination of the numerical rate of convergence.
As the final look
at the numerical results in Tables \ref{Tab.test1.E} and
\ref{Tab.test1.A}, we can see
rate of convergence of LRPI is 2.
\par
\subsection{Test case 2: SVJ models}
This example, illustrates the applicability of the proposed method to five different option pricing problem under SVJ
model with the following specifications
\begin{enumerate}
\item \emph{European put option presented in \cite{Salmi},}
\item \emph{American put option presented in \cite{Salmi},}
\item \emph{American call option with positive correlation presented in \cite{Chiarella.Ziogas, Ballestra.4},}
\item \emph{American call option with negative correlation presented in \cite{Chiarella.Ziogas, Ballestra.4},}
\item \emph{American call option presented in \cite{Toivanen.Bal,Ballestra.4},}
\end{enumerate}
where the option and model parameters are chosen as in Table \ref{Tab.par}.
\par
First, let us consider the European and American put options under stochastic volatility models with Merton's jump-diffusion considered by Salmi
el al. \cite{Salmi}. As was done in \cite{Salmi}, for the initial $S_0$ we consider three different values $S_i$ in a convenient neighborhood
of the strike $E$ so that in their the error is greater than elsewhere, i.e. $i \in \Xi_2$. Moreover we set $y_0=0.04$.
Besides, as was done in Test case 1, we suppose that $[0,S_{max}]\times[0,y_{max}]=[0,4E]\times[0,1]$.
\par
Similar to the American SV model in Test case 1, here the true price is not available (neither in the European nor in the American case), thus the true price is replaced with a reference price listed in
Tables \ref{Tab.ref.salmi.E} and \ref{Tab.ref.salmi.A}, which have been obtained in \cite{Salmi} using the projected algebraic
multigrid (PAMG) method on an extremely fine grid with 4097, 2049 and 513 nodes in $s$, $y$ and $t$ directions, respectively.
\begin{table}[tbh]\tiny
\caption{Reference prices used in European put option under SVJ model using model parameters presented in \cite{Salmi}.}
\begin{tabular*}{\columnwidth}{@{\extracolsep{\fill}}*{6}{c}}
\hline
References& & ($N_x,~N_z,~M$) &  Value at 90 & Value at 100 & Value at 110  \\
\hline
Salmi \cite{Salmi}   & & (4097,2049,513)     & 11.302917   & 6.589881   & 4.191455 \\
Proposed method   & $C^2$& (2048,1024,1024)         & 11.302908   & 6.589891   & 4.191446 \\
&                   $C^4$& (2048,1024,1024)         & 11.302912   & 6.589888   & 4.191449 \\
&                   $C^6$& (2048,1024,1024)         & 11.302913   & 6.589887   & 4.191449 \\
\hline
\end{tabular*}
\label{Tab.ref.salmi.E}
\end{table}

\begin{table}[tbh]\tiny
\caption{Reference prices used in American put option under SVJ model using model parameters presented in \cite{Salmi}.}
\begin{tabular*}{\columnwidth}{@{\extracolsep{\fill}}*{6}{c}}
\hline
References & &($N_x,~N_z,~M$) &  Value at 90 & Value at 100 & Value at 110  \\
\hline
Salmi \cite{Salmi}&    & (4097,2049,513)     & 11.619920   & 6.714240   & 4.261583 \\
Proposed method    &$C^2$& (2048,1024,1024)         & 11.619914   & 6.714249   & 4.261568 \\
&                   $C^4$& (2048,1024,1024)         & 11.619914   & 6.714247   & 4.261571 \\
&                   $C^6$& (2048,1024,1024)         & 11.619916   & 6.714247   & 4.261575 \\
\hline
\end{tabular*}
\label{Tab.ref.salmi.A}
\end{table}
\par
Applying the LRPI method based on Wendland's compactly supported radial basis functions with $C^2,~C^4$ and $C^6$ smoothness presented in this paper, we obtain the results tabulated in Tables \ref{Tab.tes2.Salmi.E.N} and \ref{Tab.tes2.Salmi.A.N}.
It is seen from the tabulated results that the approximations of option price are improved by
increasing number of sub-domains and time discretization steps. Thus, the numerical convergence of the solution is obtained.
Indeed, it can be seen that the LRPI scheme provides very fast approximation for this model.\par
Comparing the profiles of approximations, one can conclude that similar to the above discussion,
LRPI method with $C^6$ smoothness functions is more accurate than the proposed method using Wendland's radial basis functions  with lower smoothness degree, but smoothness degree of basis functions is no effect on CPU time of presented algorithm.
\begin{table}[tbh]\tiny
\caption{Test case 2: European put option under SVJ model using model parameters presented in \cite{Salmi}.}
\begin{tabular*}{\columnwidth}{@{\extracolsep{\fill}}*{11}{c}}
\hline
\multicolumn{1}{c}{}&\multicolumn{3}{c}{$C^2$}& \multicolumn{3}{c}{$C^4$}&\multicolumn{3}{c}{$C^6$}\\
\cline{2-4} \cline{5-7}\cline{8-10}
($N_x,~N_z,~M$) & RMSRD & Ratio & CPU time & RMSRD & Ratio & CPU time & RMSRD & Ratio & CPU time\\
\hline
&&&\\
(16,8,8)        & 6.72E-03 &-&0.04     & 1.08E-02 &-&0.04    & 7.20E-03 &-&0.04\\
(32,16,16)      & 3.07E-03 &1.15&0.11  & 3.50E-03 &1.62&0.11 & 2.11E-03 &1.77&0.11\\
(64,32,32)      & 1.22E-03 &1.33&0.30  & 9.51E-04 &1.88&0.30 & 8.18E-04 &1.37&0.30\\
(128,64,64)     & 2.29E-04 &2.42&0.55  & 2.53E-04 &1.91&0.55 & 2.27E-04 &1.85&0.55\\
(256,128,128)   & 6.91E-05 &1.73&1.29  & 6.59E-05 &1.94&1.29 & 6.53E-05 &1.80&1.29\\
(512,256,256)   & 1.80E-05 &1.94&4.17  & 1.67E-05 &1.98&4.17 & 1.63E-05 &2.00&4.17\\
(1024,512,512)  & 4.85E-06 &1.89&16.33 & 4.33E-06 &1.95&16.33& 4.03E-06 &2.01&16.33\\
(2048,1024,1024)& 1.59E-06 &1.61&58.46 & 1.06E-06 &2.03&58.46& 1.00E-06 &2.01&58.46\\
\hline
\end{tabular*}
\label{Tab.tes2.Salmi.E.N}
\end{table}
\begin{table}[tbh]\tiny
\caption{Test case 2: American put option under SVJ model using model parameters presented in \cite{Salmi}.}
\begin{tabular*}{\columnwidth}{@{\extracolsep{\fill}}*{11}{c}}
\hline
\multicolumn{1}{c}{}&\multicolumn{3}{c}{$C^2$}& \multicolumn{3}{c}{$C^4$}&\multicolumn{3}{c}{$C^6$}\\
\cline{2-4} \cline{5-7}\cline{8-10}
($N_x,~N_z,~M$) & RMSRD & Ratio & CPU time & RMSRD & Ratio & CPU time & RMSRD & Ratio & CPU time\\
\hline
&&&\\
(16,8,8)        & 9.91E-03 &-&0.07    & 2.55E-02 &-&0.07    & 7.48E-03 &-&0.07\\
(32,16,16)      & 1.31E-03 &2.92&0.19 & 7.68E-03 &1.73&0.19 & 2.61E-03 &1.52&0.19\\
(64,32,32)      & 4.58E-04 &1.52&0.49 & 2.19E-03 &1.81&0.49 & 7.99E-04 &1.71&0.49\\
(128,64,64)     & 1.81E-04 &1.34&0.81 & 8.20E-04 &1.42&0.81 & 3.09E-04 &1.37&0.81\\
(256,128,128)   & 7.98E-05 &1.18&1.92 & 2.26E-04 &1.86&1.92 & 8.17E-05 &1.92&1.92\\
(512,256,256)   & 2.26E-05 &1.82&6.53 & 6.16E-05 &1.88&6.53 & 2.33E-05 &1.81&6.53\\
(1024,512,512)  & 6.78E-06 &1.74&27.11& 6.99E-06 &3.14&27.11& 5.48E-06 &2.09&27.11\\
(2048,1024,1024)& 2.19E-06 &1.63&92.89& 1.76E-06 &1.99&92.89& 1.26E-06 &2.12&92.89\\
\hline
\end{tabular*}
\label{Tab.tes2.Salmi.A.N}
\end{table}
\par
To satisfy our curiosity and due to the fact that it is almost impossible to provide all cases of nodal points and time discretization steps exactly in practice, it is better to consider some of the error values using $32\times 16$, $128\times 64$ and $512\times256$ nodal points and $16$, $64$ and $256$ time discretization steps with CPU times.
\par
To clarify even more, for example, using $32\times 16$ nodal points in domain and 16 time discretization steps,
European option under SVJ model is computed with an error of
 order $O(10^{-3})$ in only $0.11$ second, instead, using $128\times 64$ nodal points and 64 time discretization steps,
 the option price is computed with an error of order $O(10^{-4})$ in $0.55$ second and using $512\times256$ nodal points and 256 time discretization steps, the option price is computed with an error of order $O(10^{-5})$ in 4.17 seconds.
On the other hand, by looking at Table \ref{Tab.tes2.Salmi.A.N}, it can also be seen that in LRPI scheme error of orders $O(10^{-3}),~O(10^{-4})$ and $O(10^{-5})$
are computed using $32\times 16$, $128\times 64$ and $512\times256$ nodal points and $16$, $64$ and $256$ time discretization steps, respectively in $0.19$, $0.81$ and $6.53$ seconds.
\par
Worthy of being considered at the end, here similar to Test case 1, the rate of convergence of LRPI is 2 (either in the European or in the American case).\\
\par
Second, we consider the same test-case presented by Chiarella et al. \cite{Chiarella.Ziogas} and Ballestra et al. \cite{Ballestra.4}, in which the model parameters and the option's data are chosen as in Table \ref{Tab.par}.
Again, we set $y_0=0.04$, $s=S_i$ so that $i \in \Xi_1$, and
$[0,S_{max}]\times[0,y_{max}]=[0,4E]\times[0,1]$.
It should be noted that the model is considered using positive and negative correlation.
The references value of option pricing is presented in Table \ref{Tab.ref.test2.ch}. In this table, the true price have been obtained in \cite{Chiarella.Ziogas}
using a finite difference approximation on an extremely fine mesh with 6000, 3000 and 1000 meshes in $s$, $y$ and $t$ directions, respectively.
\begin{table}[tbh]\tiny
\caption{Reference prices used in American call option under SVJ model using model parameters presented in \cite{Chiarella.Ziogas,Ballestra.4}.}
\begin{tabular*}{\columnwidth}{@{\extracolsep{\fill}}*{9}{c}}
\hline
&References & &($N_x,~N_z,~M$) & Value at 80 & Value at 90 & Value at 100 & Value at 110  & Value at 120\\
\hline
$\rho=0.5$ &&&&&&&\\
&Chiarella et. al. \cite{Chiarella.Ziogas}&  & (6000,3000,1000) & 1.4843   & 3.7145   & 7.7027   & 13.6722   & 21.3653\\
&Ballestra and Sgarra \cite{Ballestra.4}   & & (250,200,20)     & 1.4849   & 3.7159   & 7.7044   & 13.6735   & 21.3661\\
&Proposed method    & $C^2$&(2048,1024,1024)                         & 1.4843   & 3.7145   & 7.7027   & 13.6722   & 21.3653\\
& & $C^4$&(2048,1024,1024)                         & 1.4843   & 3.7145   & 7.7027   & 13.6722   & 21.3653\\
&    &$C^6$& (2048,1024,1024)                        & 1.4843   & 3.7145   & 7.7027   & 13.6722   & 21.3653\\
$\rho=-0.5$ &&&&&&&&\\
&Chiarella et. al. \cite{Chiarella.Ziogas}  && (6000,3000,1000) & 1.1359   & 3.3532   & 7.5970   & 13.8830   & 21.7186\\
&Ballestra and Sgarra \cite{Ballestra.4}    && (250,200,20)     & 1.1356   & 3.3537   & 7.5986   & 13.8852   & 21.7209\\
&Proposed method    &$C^2$& (2048,1024,1024)                         & 1.1359   & 3.3532   & 7.5970   & 13.8830   & 21.7186\\
&    &$C^2$& (2048,1024,1024)                         & 1.1359   & 3.3532   & 7.5970   & 13.8830   & 21.7186\\
&    &$C^2$& (2048,1024,1024)                         & 1.1359   & 3.3532   & 7.5970   & 13.8830   & 21.7186\\
\hline
\end{tabular*}
\label{Tab.ref.test2.ch}
\end{table}
\par
The results of implementing the problem by utilizing the present method with
various number of sub-domains and time discretization steps are
shown in Tables \ref{Tab.test2.ch.1} and
\ref{Tab.test2.ch.2}.
\begin{table}[tbh]\tiny
\caption{Test case 2: American call option under SVJ model using model parameters presented in \cite{Chiarella.Ziogas,Ballestra.4}, $\rho=0.5$.}
\begin{tabular*}{\columnwidth}{@{\extracolsep{\fill}}*{11}{c}}
\hline
\multicolumn{1}{c}{}&\multicolumn{3}{c}{$C^2$}& \multicolumn{3}{c}{$C^4$}&\multicolumn{3}{c}{$C^6$}\\
\cline{2-4} \cline{5-7}\cline{8-10}
($N_x,~N_z,~M$) & RMSRD & Ratio & CPU time & RMSRD & Ratio & CPU time & RMSRD & Ratio & CPU time\\
\hline
&&&&&&\\
(16,8,8)        & 2.56E-02 &-&0.07     & 5.47E-02 &-&0.07    & 5.86E-02 &-&0.07\\
(32,16,16)      & 7.10E-03 &1.85&0.19  & 1.92E-02 &1.51&0.19 & 2.48E-02 &1.24&0.19\\
(64,32,32)      & 2.36E-03 &1.59&0.49  & 5.54E-03 &1.79&0.49 & 6.74E-03 &1.88&0.49\\
(128,64,64)     & 8.52E-04 &1.47&0.81  & 1.57E-03 &1.82&0.81 & 1.77E-03 &1.93&0.81\\
(256,128,128)   & 2.30E-04 &1.89&1.92  & 4.24E-04 &1.89&1.92 & 4.54E-04 &1.96&1.92\\
(512,256,256)   & 9.67E-05 &1.25&6.53  & 1.11E-04 &1.93&6.53 & 1.12E-04 &2.02&6.53\\
(1024,512,512)  & 3.04E-05 &1.67&27.11 & 2.81E-05 &1.98&27.11& 2.67E-05 &2.07&27.11\\
(2048,1024,1024)& 9.11E-06 &1.74&92.89 & 7.07E-06 &1.99&92.89& 6.23E-06 &2.10&92.89\\
\hline
\end{tabular*}
\label{Tab.test2.ch.1}
\end{table}
\begin{table}[tbh]\tiny
\caption{Test case 2: American call option under SVJ model using model parameters presented in \cite{Chiarella.Ziogas,Ballestra.4}, $\rho=-0.5$.}
\begin{tabular*}{\columnwidth}{@{\extracolsep{\fill}}*{11}{c}}
\hline
\multicolumn{1}{c}{}&\multicolumn{3}{c}{$C^2$}& \multicolumn{3}{c}{$C^4$}&\multicolumn{3}{c}{$C^6$}\\
\cline{2-4} \cline{5-7}\cline{8-10}
($N_x,~N_z,~M$) & RMSRD & Ratio & CPU time & RMSRD & Ratio & CPU time & RMSRD & Ratio & CPU time\\
\hline
&&&&&&\\
(16,8,8)        & 4.58E-02 &-&0.07     & 6.24E-02 &-&0.07    & 1.99E-02 &-&0.07\\
(32,16,16)      & 1.63E-02 &1.49&0.19  & 1.88E-02 &1.73&0.19 & 7.82E-03 &1.35&0.19\\
(64,32,32)      & 5.70E-03 &1.52&0.49  & 5.08E-03 &1.89&0.49 & 2.14E-03 &1.87&0.49\\
(128,64,64)     & 1.79E-03 &1.67&0.81  & 1.37E-03 &1.89&0.81 & 6.31E-04 &1.76&0.81\\
(256,128,128)   & 4.74E-04 &1.92&1.92  & 3.66E-04 &1.91&1.92 & 1.60E-04 &1.98&1.92\\
(512,256,256)   & 1.42E-04 &1.74&6.53  & 9.54E-05 &1.94&6.53 & 3.97E-05 &2.01&6.53\\
(1024,512,512)  & 3.86E-05 &1.88&27.11 & 2.43E-05 &1.97&27.11& 9.86E-06 &2.01&27.11\\
(2048,1024,1024)& 1.02E-05 &1.92&92.89 & 6.16E-06 &1.98&92.89& 2.43E-06 &2.02&92.89\\
\hline
\end{tabular*}
\label{Tab.test2.ch.2}
\end{table}
\par
Overall, as already pointed out,
following the numerical findings in the different errors,
convinces us that the error of the proposed techniques decrease
very rapidly as the number of sub-domains in domain and time
discretization steps increase. In fact, for example, the price of the American option can be computed with
3 and 4 correct significants in only 0.49 and 1.92 seconds, respectively
which is excellent and very fast. We emphasize that the ratio
shown in Tables \ref{Tab.test2.ch.1} and
\ref{Tab.test2.ch.2} are second order.\\
\par
As the last experiment, we wish to find an approximation for the American call option under SVJ model considered in \cite{Toivanen.Bal}
and \cite{Ballestra.4}.
Assume that $y_0=0.04$, $s=S_i$ so that $i \in \Xi_1$, and
$[0,S_{max}]\times[0,y_{max}]=[0,4E]\times[0,1]$. The model parameters and option data are set as in Table \ref{Tab.par}.
What will be used in this model to true solution is presented in Table \ref{Tab.test2.toivan}, where they have been obtained by
performing an accurate (and also very time-consuming) simulation on a very refined mesh.
\begin{table}[tbh]\tiny
\caption{Reference prices used in American call option under SVJ model using model parameters presented in \cite{Toivanen.Bal,Ballestra.4}.}
\begin{tabular*}{\columnwidth}{@{\extracolsep{\fill}}*{8}{c}}
\hline
References && ($N_x,~N_z,~M$) & Value at 80 & Value at 90 & Value at 100 & Value at 110  & Value at 120\\
\hline
Toivanen \cite{Toivanen.Bal}  && (4096,2048,512)               & 0.328526   & 2.109397   & 6.711622   & 13.749337   & 22.143307\\
Ballestra and Sgarra \cite{Ballestra.4}    && (250,200,20)     & 0.328446   & 2.10875   & 6.711854   & 13.747836   & 22.137798\\
Proposed method  &$C^2$& (2048,1024,1024)                           & 0.328522   & 2.10939   & 6.711625   & 13.749343   & 22.143314\\
&$C^4$& (2048,1024,1024)                           & 0.328524   & 2.109393   & 6.711623   & 13.749339   & 22.143310\\
&$C^6$& (2048,1024,1024)                           & 0.328524   & 2.109394   & 6.711623   & 13.749339   & 22.143308\\
\hline
\end{tabular*}
\label{Tab.test2.toivan}
\end{table}
\par
For the sake of brevity, here we only show the Table \ref{Tab.test2.toivan.error}.
\par
\begin{table}[tbh]\tiny
\caption{Test case 2: American call option under SVJ model using model parameters presented in \cite{Toivanen.Bal,Ballestra.4}.}
\begin{tabular*}{\columnwidth}{@{\extracolsep{\fill}}*{11}{c}}
\hline
\multicolumn{1}{c}{}&\multicolumn{3}{c}{$C^2$}& \multicolumn{3}{c}{$C^4$}&\multicolumn{3}{c}{$C^6$}\\
\cline{2-4} \cline{5-7}\cline{8-10}
($N_x,~N_z,~M$) & RMSRD & Ratio & CPU time & RMSRD & Ratio & CPU time & RMSRD & Ratio & CPU time\\
\hline
&&&\\
(16,8,8)        & 1.61E-02 &-&0.07    & 2.25E-02 &-&0.07    & 5.20E-02 &-&0.07\\
(32,16,16)      & 5.61E-03 &1.52&0.19 & 8.70E-03 &1.37&0.19 & 6.41E-03 &3.02&0.19\\
(64,32,32)      & 1.89E-03 &1.57&0.49 & 2.48E-03 &1.81&0.49 & 1.88E-03 &1.77&0.49\\
(128,64,64)     & 6.25E-04 &1.60&0.81 & 6.79E-04 &1.87&0.81 & 5.41E-04 &1.80&0.81\\
(256,128,128)   & 1.87E-04 &1.74&1.92 & 1.82E-04 &1.90&1.92 & 1.48E-04 &1.87&1.92\\
(512,256,256)   & 7.24E-05 &1.37&6.53 & 4.49E-05 &2.02&6.53 & 4.08E-05 &1.86&6.53\\
(1024,512,512)  & 1.98E-05 &1.87&27.11& 1.13E-05 &1.99&27.11& 1.07E-05 &1.93&27.11\\
(2048,1024,1024)& 5.65E-06 &1.81&92.89& 2.85E-06 &1.99&92.89& 2.80E-06 &1.94&92.89\\
\hline
\end{tabular*}
\label{Tab.test2.toivan.error}
\end{table}
\par
\subsection{Test case 3: SVCJ models}
As the last test case, we aim to study the options under SVCJ model with the parameters and data which are
presented as in Table \ref{Tab.par}. Indeed, we get $y_0=0.04$, $s=S_i$ so that $i \in \Xi_2$, and
$[0,S_{max}]\times[0,y_{max}]=[0,4E]\times[0,1]$. The references prices to European and American options are tabulated
in Table \ref{Tab.test3}, which have been obtained in \cite{Salmi} using the projected algebraic
multigrid (PAMG) method on an extremely fine grid with 4097, 2049 and 513 nodes in $s$, $y$ and $t$ directions, respectively.
\begin{table}[tbh]\tiny
\caption{Reference prices under SVCJ model using model parameters presented in Test case 3.}
\begin{tabular*}{\columnwidth}{@{\extracolsep{\fill}}*{7}{c}}
\hline
&References && ($N_x,~N_z,~M$) & Value at 90 & Value at 100 & Value at 110  \\
\hline
European option &&&&&&\\
&Salmi \cite{Salmi}  && (4097,2049,513) & 11.134438   & 6.609162   & 4.342956\\
&Proposed method &$C^2$ & (2048,1024,1024)   & 11.134424   & 6.609173   & 4.342972\\
&                &$C^4$ & (2048,1024,1024)   & 11.134429   & 6.609169   & 4.342967\\
&                &$C^6$ & (2048,1024,1024)   & 11.134433   & 6.609167   & 4.342964\\
American option &&&&&&\\
&Salmi \cite{Salmi}  && (4097,2049,513) & 11.561620   & 6.780527   & 4.442032 \\
&Proposed method  &$C^2$& (2048,1024,1024)   & 11.561601   & 6.780541   & 4.442047 \\
&                 &$C^4$& (2048,1024,1024)   & 11.561611   & 6.780536   & 4.442044 \\
&                 &$C^6$& (2048,1024,1024)   & 11.561615   & 6.780533   & 4.442042 \\
\hline
\end{tabular*}
\label{Tab.test3}
\end{table}
\par
Implementing the proposed numerical technique produce the outcomes given by Tables \ref{Tab.test3.E} and \ref{Tab.test3.A}.
These tables show the numerically identified solutions for option pricing in European and American option models, respectively.
\begin{table}[tbh]\tiny
\caption{Test case 3: European option.}
\begin{tabular*}{\columnwidth}{@{\extracolsep{\fill}}*{11}{c}}
\hline
\multicolumn{1}{c}{}&\multicolumn{3}{c}{$C^2$}& \multicolumn{3}{c}{$C^4$}&\multicolumn{3}{c}{$C^6$}\\
\cline{2-4} \cline{5-7}\cline{8-10}
($N_x,~N_z,~M$) & RMSRD & Ratio & CPU time & RMSRD & Ratio & CPU time & RMSRD & Ratio & CPU time\\
\hline
&&&&&&\\
(16,8,8)        & 3.41E-02 &-&0.13      & 1.59E-02 &-&0.13    & 7.86E-03 &-&0.13\\
(32,16,16)      & 5.59E-03 &2.61&0.22   & 4.93E-03 &1.69&0.22 & 3.15E-03 &1.32&0.22\\
(64,32,32)      & 1.53E-03 &1.87&0.71   & 1.35E-03 &1.87&0.71 & 9.30E-04 &1.76&0.71\\
(128,64,64)     & 4.05E-04 &1.92&1.81   & 3.68E-04 &1.88&1.81 & 2.58E-04 &1.85&1.81\\
(256,128,128)   & 1.42E-04 &1.51&2.49   & 9.66E-05 &1.93&2.49 & 7.73E-05 &1.74&2.49\\
(512,256,256)   & 3.64E-05 &1.96&6.02   & 2.50E-05 &1.95&6.02 & 1.96E-05 &1.98&6.02\\
(1024,512,512)  & 9.30E-06 &1.97&24.53  & 6.42E-06 &1.96&24.53& 4.86E-06 &2.01&24.53\\
(2048,1024,1024)& 2.44E-06 &1.93&83.41  & 1.65E-06 &1.96&83.41& 1.18E-06 &2.04&83.41\\
\hline
\end{tabular*}
\label{Tab.test3.E}
\end{table}
\begin{table}[tbh]\tiny
\caption{Test case 3: American option.}
\begin{tabular*}{\columnwidth}{@{\extracolsep{\fill}}*{11}{c}}
\hline
\multicolumn{1}{c}{}&\multicolumn{3}{c}{$C^2$}& \multicolumn{3}{c}{$C^4$}&\multicolumn{3}{c}{$C^6$}\\
\cline{2-4} \cline{5-7}\cline{8-10}
($N_x,~N_z,~M$) & RMSRD & Ratio & CPU time & RMSRD & Ratio & CPU time & RMSRD & Ratio & CPU time\\
\hline
&&&&&&\\
(16,8,8)        & 1.51E-02 &-&0.13     & 1.63E-02 &-&0.20      & 1.11E-02 &-&0.20\\
(32,16,16)      & 2.66E-03 &2.51&0.22  & 2.40E-03 &2.76&0.38   & 2.41E-03 &2.21&0.38\\
(64,32,32)      & 6.97E-04 &1.93&0.71  & 6.31E-04 &1.93&1.01   & 8.89E-04 &1.44&1.01\\
(128,64,64)     & 2.97E-04 &1.23&1.81  & 2.08E-04 &1.60&3.62   & 2.50E-04 &1.83&3.62\\
(256,128,128)   & 8.65E-05 &1.78&2.49  & 6.72E-05 &1.63&4.17   & 7.71E-05 &1.70&4.17\\
(512,256,256)   & 2.59E-05 &1.74&6.02  & 2.14E-05 &1.65&10.96  & 2.08E-05 &1.89&10.96\\
(1024,512,512)  & 8.42E-06 &1.62&24.53 & 6.32E-06 &1.76&40.00  & 5.45E-06 &1.93&40.00\\
(2048,1024,1024)& 2.47E-06 &1.77&83.41 & 1.79E-06 &1.82&137.21 & 1.42E-06 &1.94&137.21\\
\hline
\end{tabular*}
\label{Tab.test3.A}
\end{table}
\par
It is seen from Tables \ref{Tab.test3.E} and \ref{Tab.test3.A} that the approximations are improved by increasing the number of nodes and time
discretization steps, since the RMSRD values tend to zero more quickly as the node distributions increase, which confirms convergence property of the proposed method agian. Once again, we wish to state the numerical rate of convergence in the context of this test case.
It can be seen that similar to the SV and SVJ models in two previous test cases, we obtain Ratio=2 by applying the LRPI method based on Wendland's compactly supported radial basis functions with $C^2,~C^4$ and $C^6$ smoothness. The approximations are exhibited in Tables \ref{Tab.test3.E} and \ref{Tab.test3.A} to verify this fact. Putting all these things together, we conclude that the numerical
methods proposed in this paper are accurate, convergence and fast.
\par
\begin{figure}
\includegraphics[width=6in]{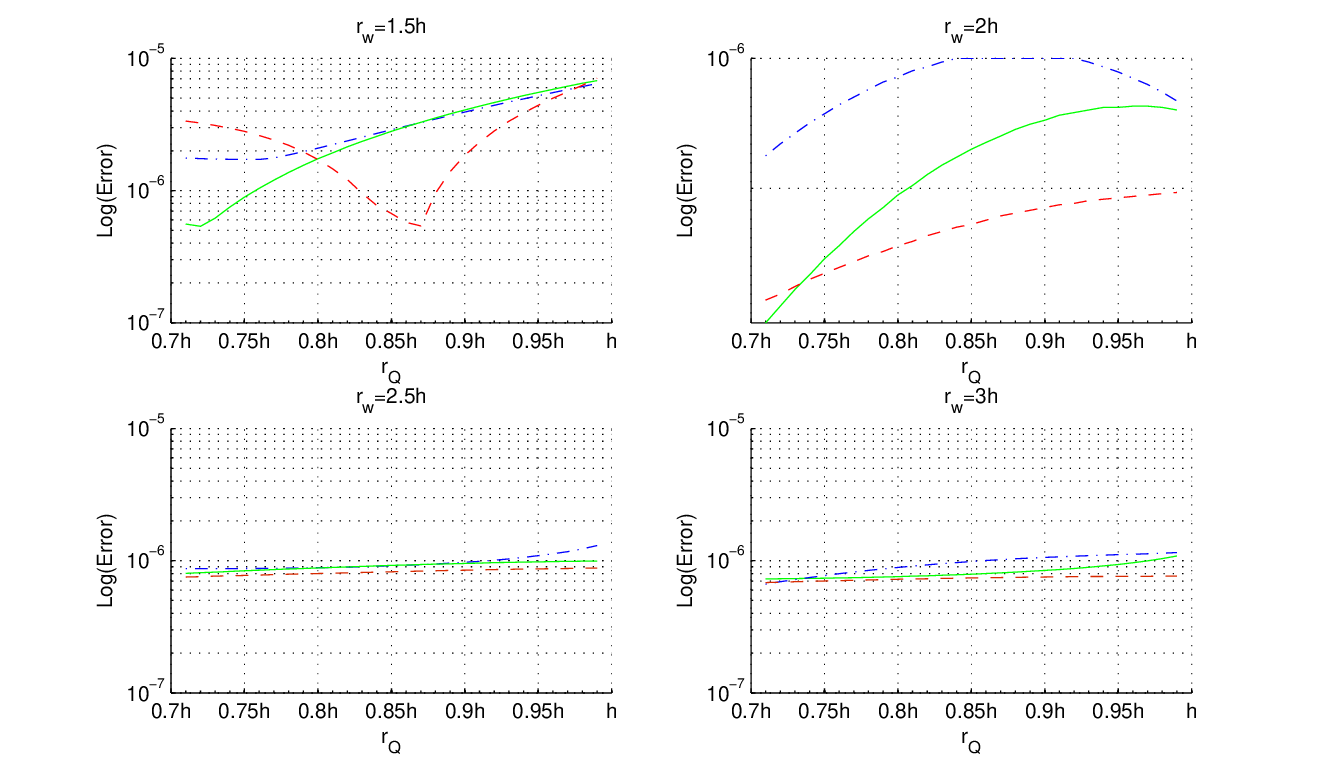}
\caption{Test1: European option.}
\label{Test1E}
\end{figure}
\clearpage
\begin{figure}
\includegraphics[width=6in]{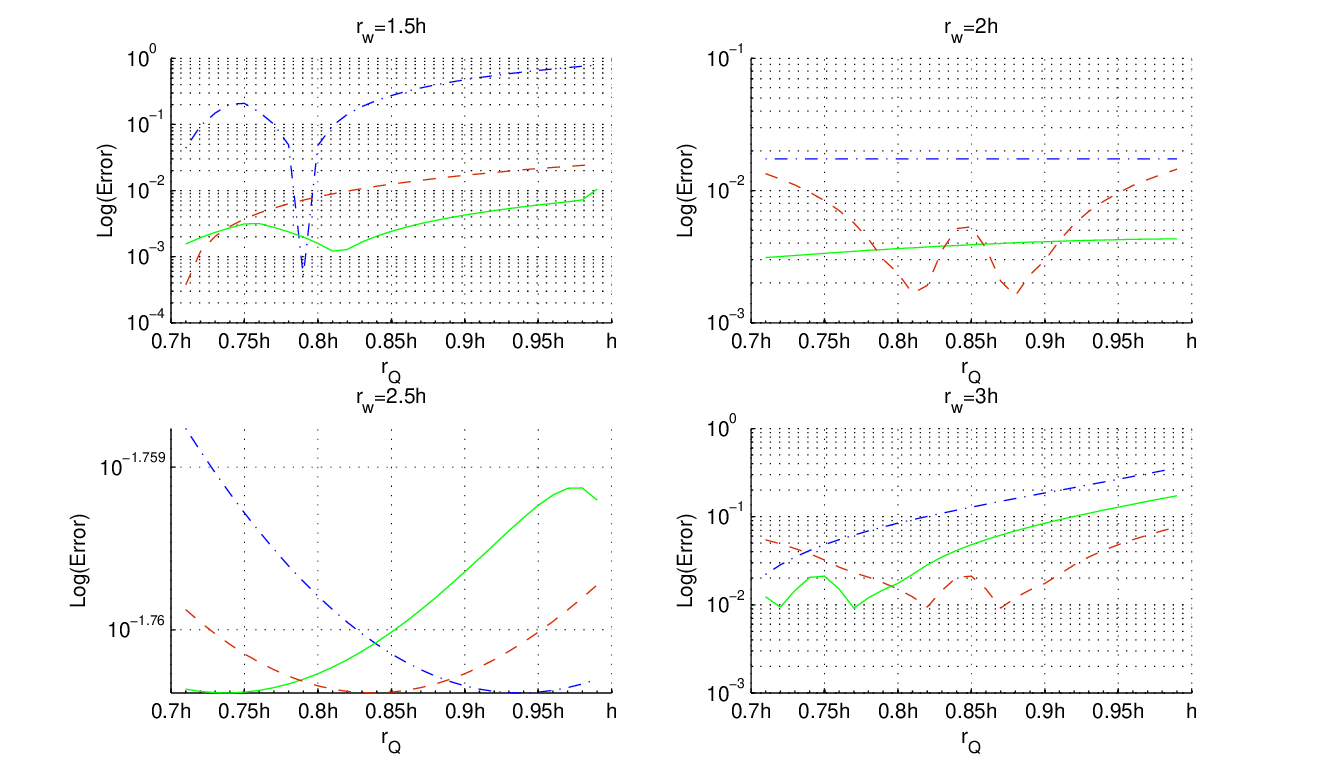}
\caption{Test1: American option.}
\label{Test1A}
\end{figure}
\begin{figure}
\includegraphics[width=6in]{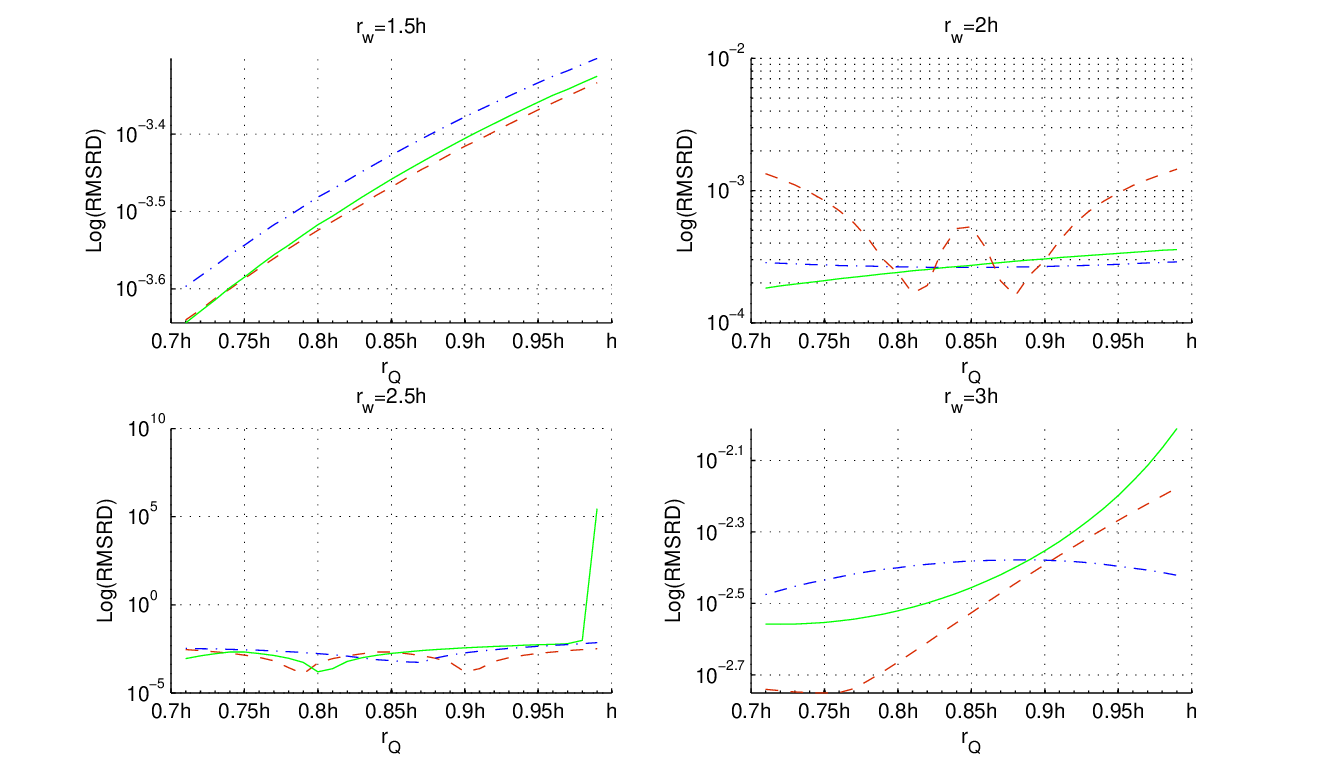}
\caption{Test case 2: European put option under SVJ model using model parameters presented in \cite{Salmi}.}
\label{test2E}
\end{figure}
\begin{figure}
\includegraphics[width=6in]{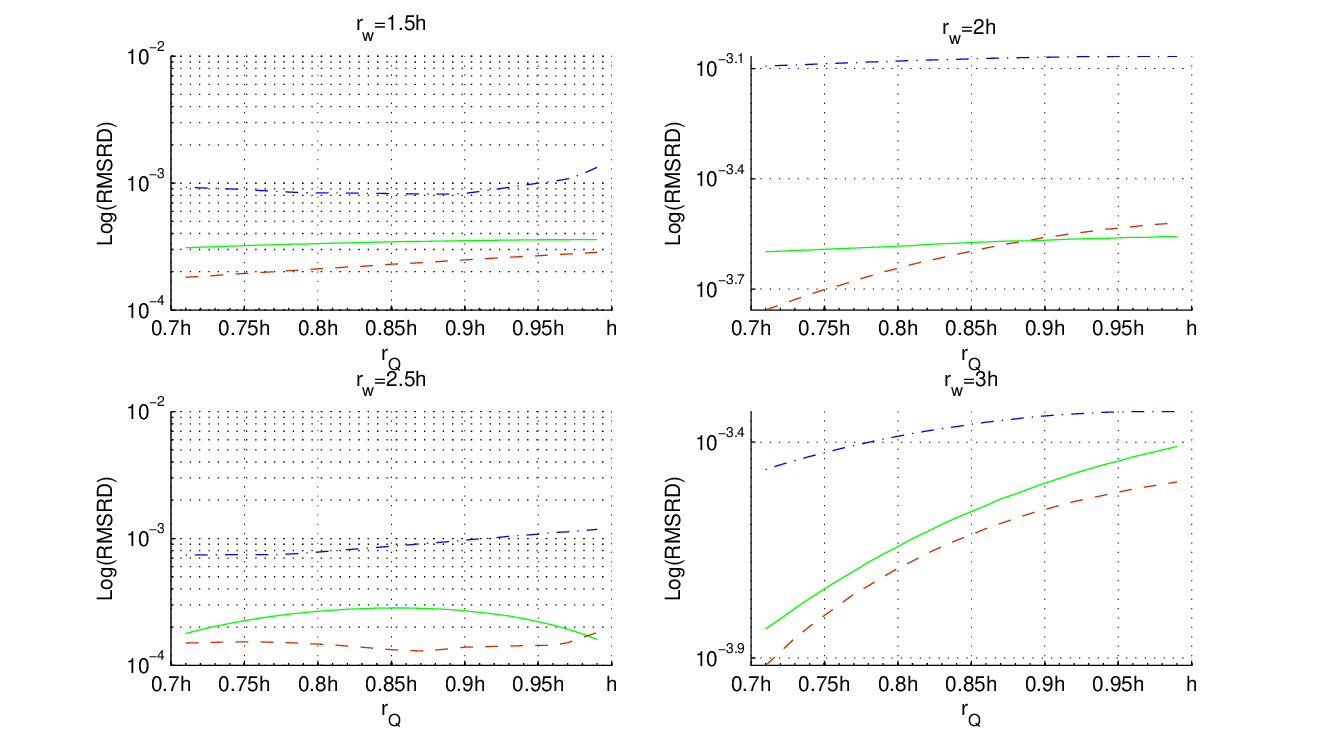}
\caption{Test case 2: American put option under SVJ model using model parameters presented in \cite{Salmi}.}
\label{test2Aa}
\end{figure}
\begin{figure}
\includegraphics[width=6in]{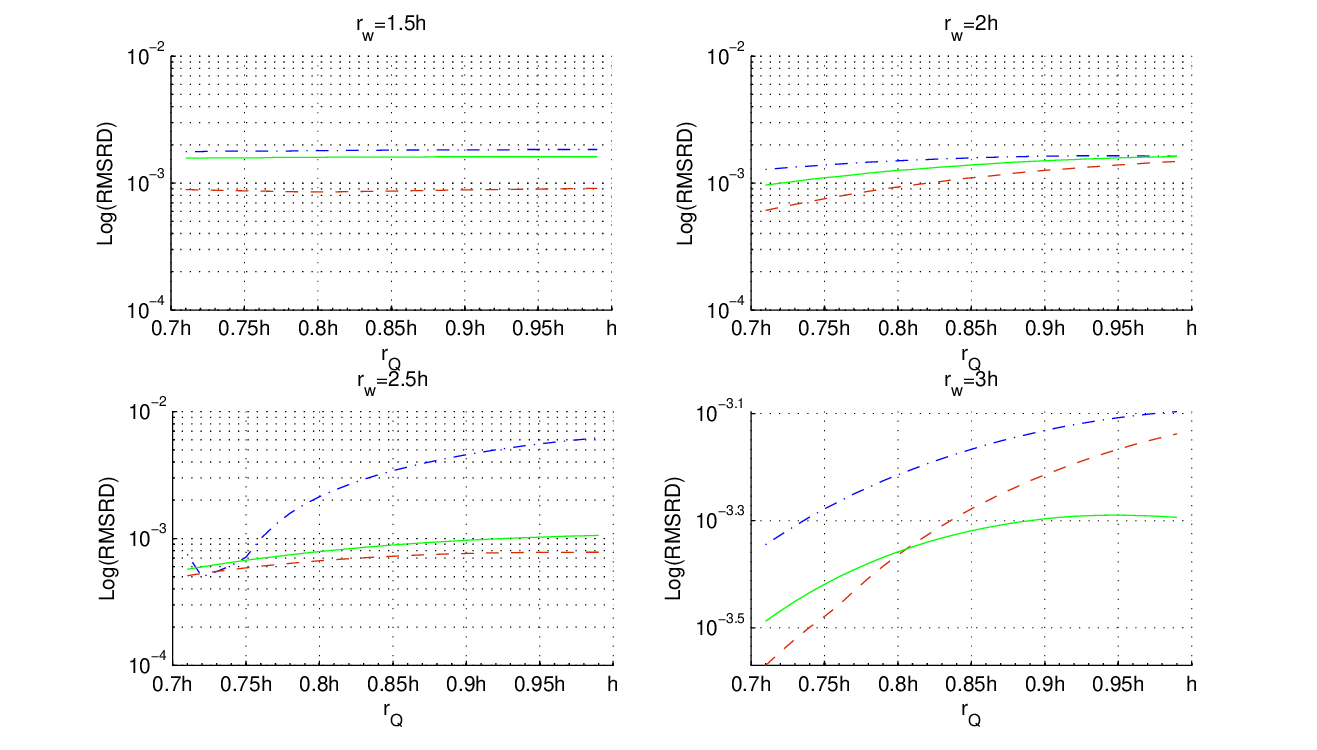}
\caption{Test case 2: American call option under SVJ model using model parameters presented in \cite{Chiarella.Ziogas,Ballestra.4}, $\rho=0.5$.}
\label{test2Ab}
\end{figure}
\begin{figure}
\includegraphics[width=6in]{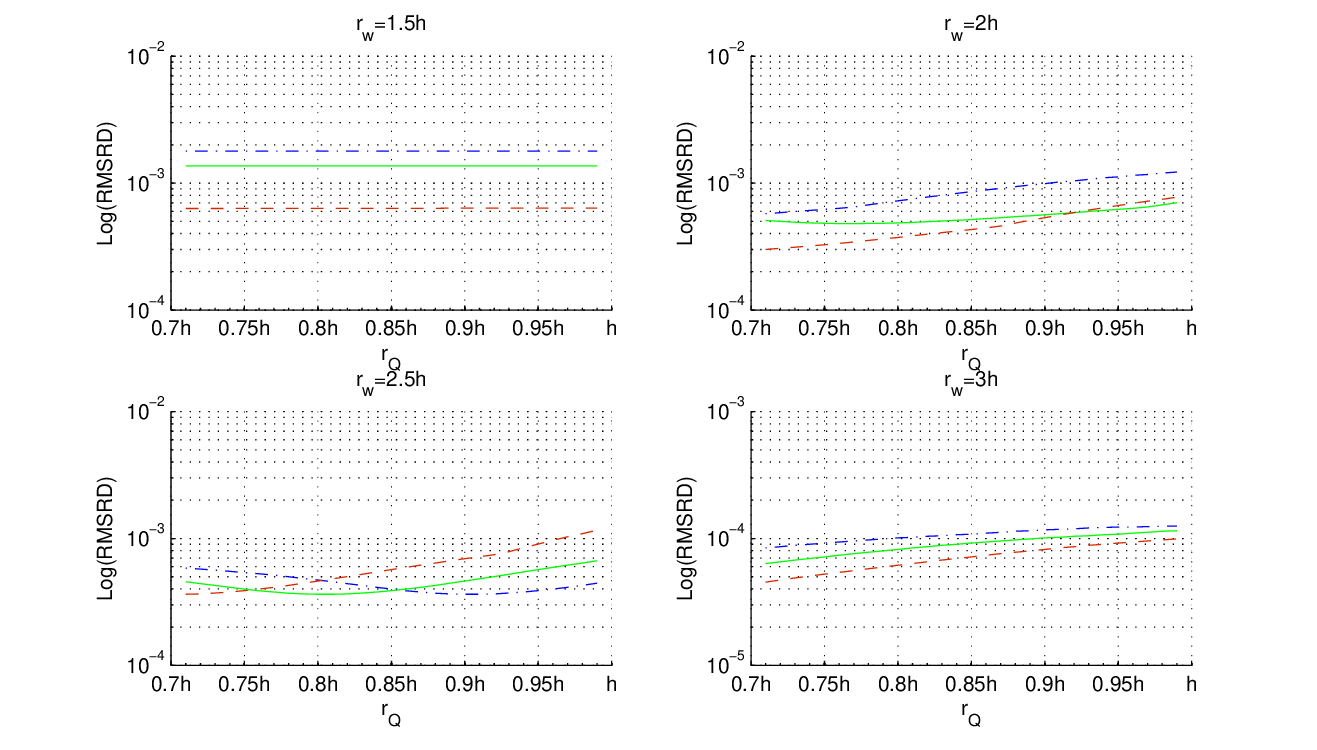}
\caption{Test case 2: American call option under SVJ model using model parameters presented in \cite{Chiarella.Ziogas,Ballestra.4}, $\rho=-0.5$.}
\label{test2Ac}
\end{figure}
\begin{figure}
\includegraphics[width=6in]{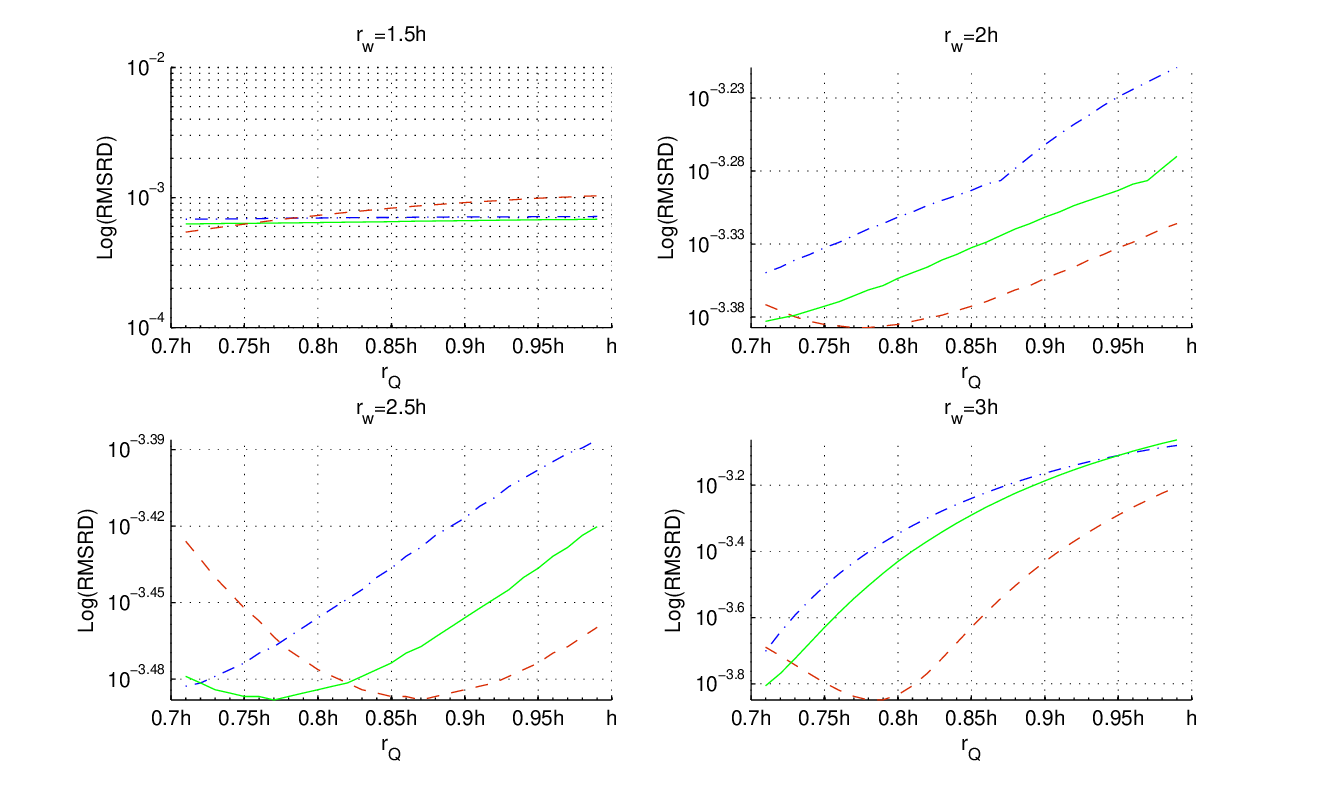}
\caption{Test case 2: American call option under SVJ model using model parameters presented in \cite{Toivanen.Bal,Ballestra.4}.}
\label{test2Ad}
\end{figure}
\begin{figure}
\includegraphics[width=6in]{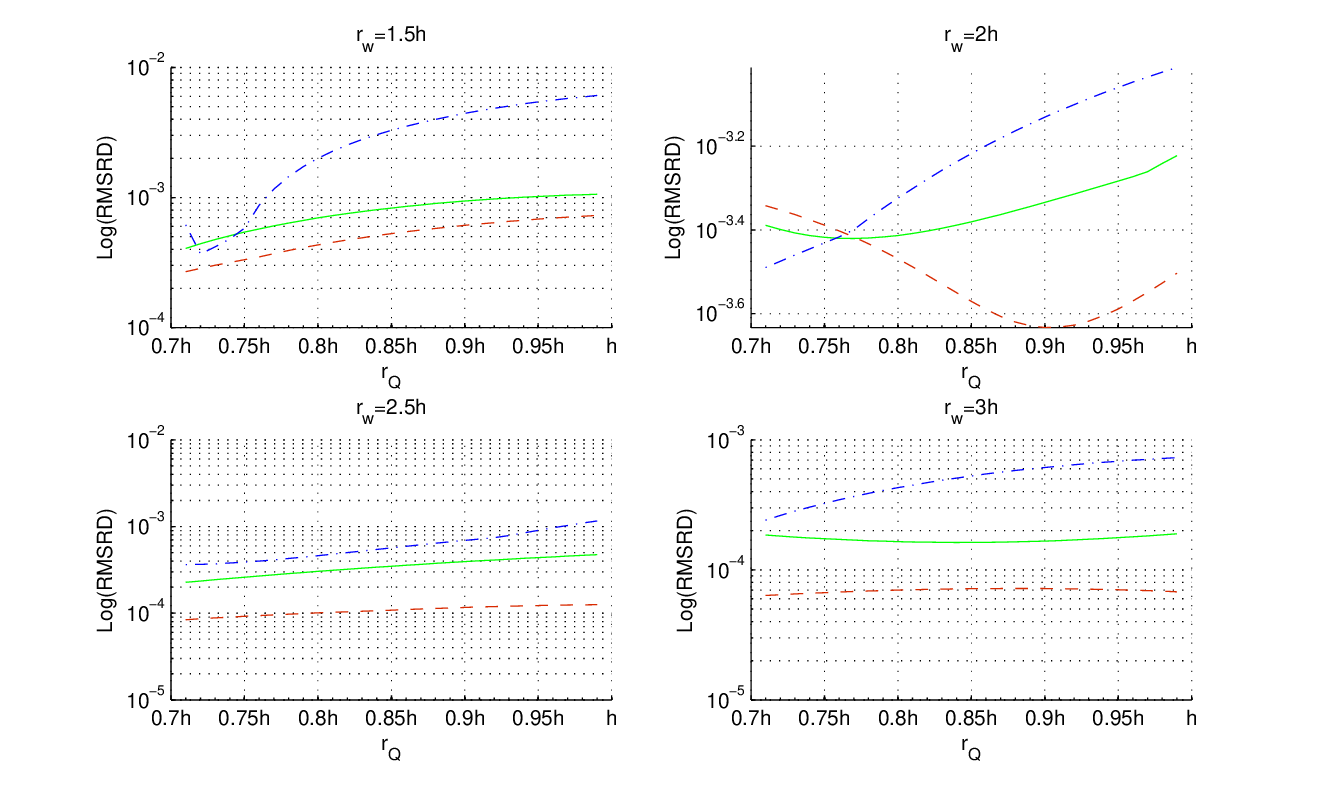}
\caption{Test case 3: European option.}
\label{test3E}
\end{figure}
\begin{figure}
\includegraphics[width=6in]{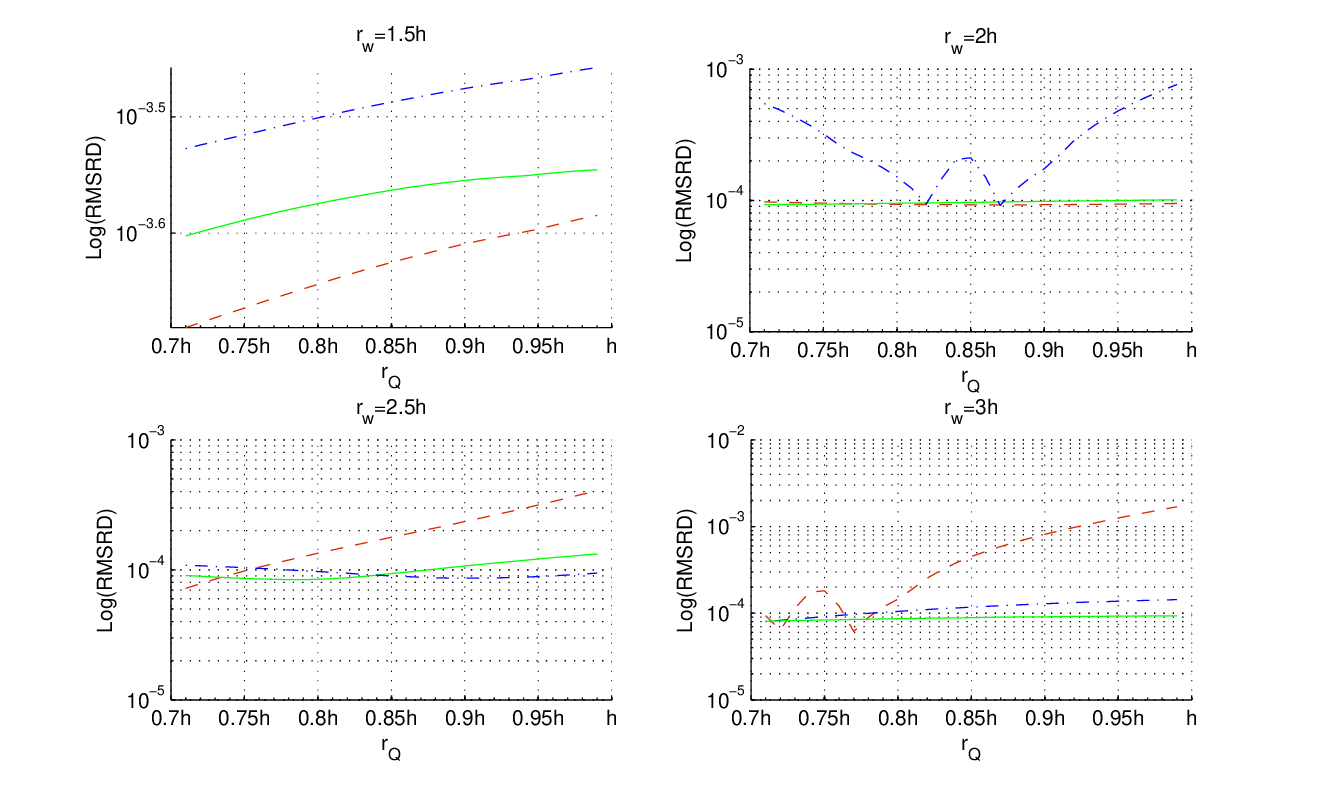}
\caption{Test case 3: American option.}
\label{test3A}
\end{figure}

\section{Conclusions}
During the last decade, many works have been done to find modifications of classical Black-Scholes model to satisfy these phenomena in financial markets
such as the models with stochastic volatility (SV), stochastic volatility models with jumps (SVJ), and stochastic volatility
models with jumps in returns and volatility (SVCJ).
Hence, an analytical solution for pricing these options is impossible. Therefore, to solve these problems, we need to have a powerful computational method.
For the first time in mathematical financial field, we proposed the
local weak form meshless methods for option pricing under SV, SVJ and SVCJ models; especially in
this paper we focused on one of their scheme named local radial point interpolation
(LRPI), based on Wendland's compactly supported radial basis functions
(WCS-RBFs) with $C^6$, $C^4$ and $C^2$ smoothness degrees.
\par
Overall the numerical achievements that should be highlighted here
are as follows:\par (1) The price of American
option is computed by Richardson extrapolation of the price of
Bermudan option. In essence the Richardson extrapolation reduces
the free boundary problem and linear complementarity problem to a
fixed boundary problem, which is much simpler to solve. Thus,
instead of describing the aforementioned linear complementarity
problem or penalty method, we directly focus our attention on
the partial differential equation satisfied by the price of a
Bermudan option which is faster and more accurate than other
methods.
\par
(2) The infinite space domain $\mathbb{R}^+\times \mathbb{R}^+$ is truncated to $[0, S_{max}]\times[0,y_{max}]$ in SVJ and SVCJ models, with the
sufficiently large values $S_{max}$ and $y_{max}$ to avoid an unacceptably large truncation error.
The options' payoffs considered in this paper are non-smooth functions, in particular
 their derivatives, are discontinuous at the strike price. Therefore, to reduce as much as possible the losses of accuracy, the points of the trial functions are concentrated on a spatial region close to the strike prices. So, we employ the change of variables proposed by Clarke and Parrott \cite{Clarke.Parrott}.
\par
(3) As far as the time discretization is concerned, we used the implicit-explicit (IMEX) time stepping scheme, which is unconditionally
stable and allows us to smooth the discontinuities of the options' payoffs.
Note that, in stochastic volatility model with jumps, the integral part is a non-local integral, whereas the other parts which are differential operators, are all local.
No doubt, since the integral part is non-local operator, a dense linear system of equations will be obtained by using the $\theta$-weighted discretization scheme.
Therefore, to obtain a sparse linear system of equations, it is better to use an IMEX scheme which is noted for avoiding dense matrices.
So far, and to the best of our knowledge, published work existing in the literature which use the IMEX scheme to price the options, include \cite{Salmi}.
Such an approach is only first-order
accurate, however a second-order time discretization is obtained by performing a Richardson extrapolation
procedure with halved time step.
\par
(4) Stability analysis of the method is analyzed and performed by the matrix method in the present paper.
\par
(5) Up to now, only strong form meshless methods based on radial
basis functions (RBFs) have been used for option
pricing under SV model \cite{Ballestra.1}. These techniques yield
high levels of accuracy, but have of a very serious drawback such
as produce a very ill-conditioned systems and very sensitive to
the select of collocation points.
Again, we do emphasize that in the new methods presented in this manuscript, coefficient matrix of the linear systems are sparse.
\par
(6) LRPI scheme is the truly meshless methods,
because, a traditional non-overlapping, continuous mesh is not
required, neither for the construction of the shape functions, nor
for the integration of the local sub-domains.
\par
(7) Meshless methods using global RBFs such as Gaussian and multiquadric RBFs have a free parameter known as shape parameter.
Despite many research works which are done to find
algorithms for selecting the optimum values of $\epsilon$
\cite{Cheng.Golberg.Kansa.Zammito,Carlson.Foley,G.E.Fasshauer.J.G.Zhang,S.Rippa,A.E.Tarwater}, the optimal choice of shape parameter is an open
problem which is still under intensive investigation.
In general, as the value of the shape parameter $\epsilon$ decreases, the matrix
of the system to be solved becomes highly ill-conditioned.
To overcome this drawback of the global RBFs, the local RBFs such as Wendland compactly supported radial
basis functions, which are local and stable functions, are proposed which are applied in this work.
\par
(8) In LRPI method,
using the delta Kronecker property, the boundary conditions can be
easily imposed.
\par
(9) The optimal values of the size
of local sub-domain and support domain ($r_Q$ and $r_w$) are illustrated using the Figures \ref{Test1E} and \ref{Test1A} for SV model; Figures \ref{test2E}, \ref{test2Aa}, \ref{test2Ab}, \ref{test2Ac} and \ref{test2Ad} for SVJ model; and Figures \ref{test3E} and \ref{test3A} for SVCJ model
for error vs. different values of $r_Q$ and $r_w$.
\par
(10) Numerical
experiments are presented showing that the LBIE and LRPI
approaches are extremely accurate and fast.
\par
(11) Future work will concern an extension
to the basket options.

\clearpage\textbf{References}
\bibliographystyle{elsarticle-num}
\bibliography{hestonjump}
\end{document}